\documentclass[nofootinbib,floatfix,superscriptaddress,twocolumn]{revtex4}
\usepackage{graphicx}
\usepackage{epsfig}
\usepackage{bm}
\usepackage[T1]{fontenc}
\usepackage[latin9]{inputenc}
\usepackage{amssymb}
\usepackage{float}
\usepackage{amsmath}
\usepackage{dcolumn}
\usepackage{cancel}
\usepackage[colorlinks]{hyperref}
\usepackage[usenames,dvipsnames]{color}
\hypersetup{
     breaklinks=true,
    pdfstartview={FitH},    
    colorlinks=true,       
    linkcolor=blue,          
    citecolor=red,        
    filecolor=magenta,      
    urlcolor=blue,           
    anchorcolor=green,      
    linktocpage=true
}

\newcommand{\be}{\begin{equation}}
\newcommand{\ee}{\end{equation}}

\begin{document}

\title{Black hole geometrothermodynamics and critical phenomena: a look from Tsallis entropy-based perspective}

\author{G.~G.~Luciano}
\email{giuseppegaetano.luciano@udl.cat}
\affiliation{Applied Physics Section of Environmental Science Department,  Escola Polit\`ecnica Superior, Universitat de Lleida, Av. Jaume
II, 69, 25001 Lleida, Spain}

\author{A. Sheykhi}
\email{asheykhi@shirazu.ac.ir}
\affiliation{Department of Physics, College of Sciences, Shiraz University, Shiraz 71454, Iran}
\affiliation{Biruni Observatory, College of Sciences, Shiraz University, Shiraz 71454, Iran}

\date{\today}

\begin{abstract}
We analyze geometrothermodynamics of charged anti-de Sitter (AdS) black holes with a global monopole in the framework of Tsallis
statistics. The latter arises from a non-additive generalization
of Boltzmann-Gibbs entropy, which is still recovered as a
particular sub-case. We examine the effects of Tsallis
entropy on small-large black holes phase-transitions
and critical exponents, comparing the results with the
liquid-gas change of phase occurring in van der Waals fluids. We
also discuss some physical constraints on Tsallis parameter and the
impact on sparsity of BH radiation. Thermodynamic curvature behavior is then explored from a geometric perspective within Ruppeiner formalism on the $S-P$ coordinate space. Similar considerations are extended to charged $(2+1)$-dimensional Banados-Teitelboim-Zanelli (BTZ) black holes by computing the entropy-pressure and entropy-volume corrected curvatures. Our analysis shows that Tsallis prescription has non-trivial implications on black hole geometrothermodynamics, emphasizing the role of non-additive entropy on the correspondence between black holes and van der Waals systems.
\end{abstract}

 \maketitle

\section{Introduction}
\label{Intro} Since the seminal works of
Bekenstein~\cite{Bekenstein:1973ur,Bekenstein:1974ax},
Hawking~\cite{Hawking:1975vcx,Hawking:1976de} and later
developments~\cite{Bardeen:1973gs,Gibbons:1976ue}, the thermodynamics of black holes (BHs) has been largely explored, also in view of its potential impact on understanding quantum gravity. In parallel, the observation that asymptotically Anti-de Sitter (AdS) BHs admit a gauge duality description in the language of dual
thermal field theory has motivated a fluid-like modeling of
the underlying degrees of freedom~\cite{Chamblin:1999tk,Chamblinbis}, emphasizing the correspondence between BHs and
condensed matter systems and catalyzing the attention on 
the study of BH thermodynamics through geometrical techniques - geometrothermodynamics ~\cite{Davies:1977bgr,Cai:1998ep,Quevedo:2007mj,Quevedo:2008ry,Sahay:2010tx,Wei:2015iwa,Dehyadegari:2016nkd,Li:2017xvi,Wei:2019uqg,Xu:2020gud,KordZangeneh:2017lgs,Wei:2019yvs,Ghosh:2019pwy}. 

In this picture, the standard way to analyze
the nature of the microscopic interactions among BH constituents
is by considering their macroscopic geometrothermodynamic properties. For instance, in the case of Weinhold~\cite{Wein1} and Ruppeiner~\cite{Rupp1,Rupp2,Rupp3} formalisms, the sign of the  scalar curvature of the metric reveals the global behavior of the intermolecular interactions, with negative curvature corresponding to attractive forces and vice-versa. Also, flat metric characterizes BHs with perfectly balanced interactions. The microscopic behavior of BHs in geometrothermodynamics has been examined for a vast class of systems, such as $(2+1)$-dimensional Banados-Teitelboim-Zanelli (BTZ) BHs~\cite{Cai:1998ep}, Reissner-Nordstr\text{\"o}m, Kerr and
Reissner-Nordstr\text{\"o}m-AdS
BHs~\cite{Wei:2015iwa,Wei:2019uqg}, Schwarzschild-AdS
BHs~\cite{Xu:2020gud} and exotic BTZ BHs~\cite{Ghosh:2020kba},
with different results on the attractive/repulsive nature of
related forces.

A crucial ingredient in BH thermodynamics is the definition of
entropy. Physically meaningful arguments (including, for example, the holographic principle~\cite{tHooft:1993dmi,Susskind:1994vu}) exist that the entropy of a three-dimensional BH may scale as the surface area $A_{bh}$ of its boundary rather than volume $V$ according to the Bekenstein-Hawking entropy-area law $S_{BH}=A_{bh}/A_0$, where $A_0=4$ is the Planck area (here and henceforth we adopt geometric units $\hslash=c=k_B=G=1$).
However, there is quite widespread recognition that Bekenstein-Hawking entropy is somehow unconventional, as it violates thermodynamic extensivity. Indeed,
should BHs be considered as  genuine two-dimensional systems, then $S_{BH}$ could be properly identified with the thermodynamic
entropy. On the other hand, if we regard BHs as three-dimensional
systems - as arguably more natural - $S_{BH}$ cannot be taken
as a correct definition anymore. 

The above issue has been addressed
in~\cite{TsallisCirto}, showing that a non-additive entropic
functional is to be introduced for BHs and, in general, for
complex systems exhibiting long-range interactions (like
gravitational systems) and/or long-time memory. The resulting deformed entropy reads\footnote{The modified entropy~\eqref{TsEn} is known as Tsallis-Cirto entropy~\cite{TsallisCirto}. For brevity,   
we shall simply name it Tsallis entropy in the following.}
\begin{equation}
\label{TsEn}
S_{\delta}= \left(\frac{A_{bh}}{A_0}\right)^{\delta}\,,
\end{equation}
where $\delta$ quantifies deviation from additivity, in such a way that $S_\delta\rightarrow S_{BH}$ for $\delta\rightarrow1$.  Modified Friedmann equations
and their cosmological consequences, based on Tsallis entropy
(\ref{TsEn}), have been investigated in \cite{Sheykhi1,Saridakis:2018unr,LambGhosh,SaridakisLympEPJC,LambJiz}. Employing the
non-additive Tsallis entropy for the large-scale gravitational
systems, it has been argued that modified Friedmann equations
admit an accelerated expansion, without invoking any kind of dark
energy, provided one chooses $\delta<1/2$. In the non-relativistic
regime, however, if one takes $\delta \lesssim1/2$, the modified
Newton's law of gravitation can explain the flat galactic rotation
curves without invoking particle dark matter \cite{Sheykhi2}.
In passing, we mention that
constraints on $|\delta-1|$ have been set in different physical
scenarios, ranging from BH physics, to cosmology and quantum field
theory (see~\cite{Luciano:2022ely} for a review). Although not
contemplated in the original formulation, some studies have also
allowed $\delta$ to be varying over time/energy
scale~\cite{App13,Lucianomix,Lucianomix2,Jizba:2022icu}. Recently,
BH thermodynamics based on Eq.~\eqref{TsEn} has been explored in connection with generalized models of Heisenberg uncertainty
principle~\cite{Alonso,Cimidiker:2023kle}.

Thermodynamics of BHs has some apparent pitfalls
comparing to that of ordinary matter. It is known that 
the thermodynamic description 
of some classes of BHs depends on the ensemble where they are investigated~\cite{Chamblin:1999tk}. 
Furthermore, the definition of thermodynamic 
volume turns out to be subtle, especially in the presence of a negative cosmological constant. In this context, the analysis of~\cite{Kubiznak} has involved a prevailing proposal~\cite{Kastor:2009wy,Cvetic:2010jb}, which envisages that the cosmology constant
and its conjugate quantity should be identified with BH thermodynamic pressure and volume, respectively, and properly included in the first law of BH thermodynamics. 
Once the variation of $\Lambda$ is included in the first
law, the black hole mass is identified with enthalpy rather than internal energy~\cite{Kastor:2009wy}. In the ensuing extended phase space, AdS BHs have been found to exhibit thermodynamic phase transitions that resemble in many aspects the changes of phase occurring in van der Waals fluids~\cite{Chamblin:1999tk,Altamirano:2014tva}. The most suggestive example is provided by the first-order phase transition between Schwarzschild AdS BHs and 
thermal AdS background described by Hawking and Page in~\cite{Hawking:1982dh}  (Hawking-Page phase transition).

Phase transitions in the evolution of the
early Universe can create topological defects like global
monopoles~\cite{Kibble:1976sj,Vilenkin:1984ib}. Gravitational
fields associated with these solutions generate fluctuations in
the microwave background radiation, which later evolve into the
macro scale of galaxy clusters and are potentially observable.
As shown in~\cite{Vilenkin:2000jqa}, global monopoles originate from
breaking of a gauge symmetry which carries a unit of magnetic
flux. In~\cite{Barriola:1989hx} the existence of monopoles has
been originally traced back to the breakdown of the global SO(3) 
symmetry of the scalar triplet field in Schwarzschild BH geometry
into U(1) symmetry. Since then, the physical properties of BHs with global monopoles have been investigated
extensively~\cite{Jing:1993np,Yu:1994fy,Li:2002ku,Jiang:2005xb,Carames:2017ngt}
and a connection with superconducting~\cite{Chen:2009vz} and van
der Waals-like~\cite{Deng:2018wrd} phase transitions has been
established. However, to the best of our knowledge, a systematic
study of phase transitions of BHs with global monopoles in the
most suited Tsallis entropy-based thermodynamics has not been conducted yet.

The aim of the present work is to fill this gap. 
In more detail, we analyze the effects of
Tsallis entropy~\eqref{TsEn} 
on geometrothermodynamics and phase transitions of
charged AdS BHs with global monopole. The importance of such BHs lies in the fact that the thermodynamically stable BHs exist only in AdS space. Furthermore, they provide a unique arena to conduct research on the gravity analogue systems~\cite{Malda,Iorio}. In some sense, the highlight of this study not only improves our knowledge of BH thermodynamics, but potentially gives better insights to related issues straddling both the classical and quantum aspects of gravity. 
Investigation proceeds as
follows: in the next Section, we review basic notions of van
der Waals fluids, which are useful for subsequent analogy with BHs.
For this purpose, we mostly 
follow~\cite{Kubiznak}. In Sec.~\ref{Pht}  we
address phase transitions of charged AdS BHs
with global monopole in the framework of 
Tsallis statistics. Sec.~\ref{Geom} is devoted to study
geometrothermodynamics of AdS BHs. We also report on the case of
BTZ BHs for comparison with literature. Conclusions and outlook
are finally discussed in Sec.~\ref{Conc}.

\section{Van der Waals fluid}
\label{vdwSec}

Van der Waals model provides the simplest description of
the behavior of real fluids made of finite-size interacting
molecules. It also allows us to understand the qualitative features
of the liquid-gas phase transition. The characteristic equation of
van der Waals fluid is 
\be 
\label{vdwequation}
\left(P+\frac{a}{v^2}\right)\left(v-b\right) = T\,, 
\ee 
where $v=V/N$ is the specific volume of the fluid and $N$, $P$, $T$ the number of its constituents, pressure and temperature,
respectively.  The constant $a>0$ is related to the attractive
interaction among the molecules of the fluid, while $b>0$ to their
finite size. Van der Waals equation can be equivalently cast as \be \label{pv}
Pv^3-\left(T+bP\right)v^2+a\left(v-b\right)=0\,. 
\ee

\begin{figure}[t]
\begin{center}
\hspace{-3mm}\includegraphics[width=8cm]{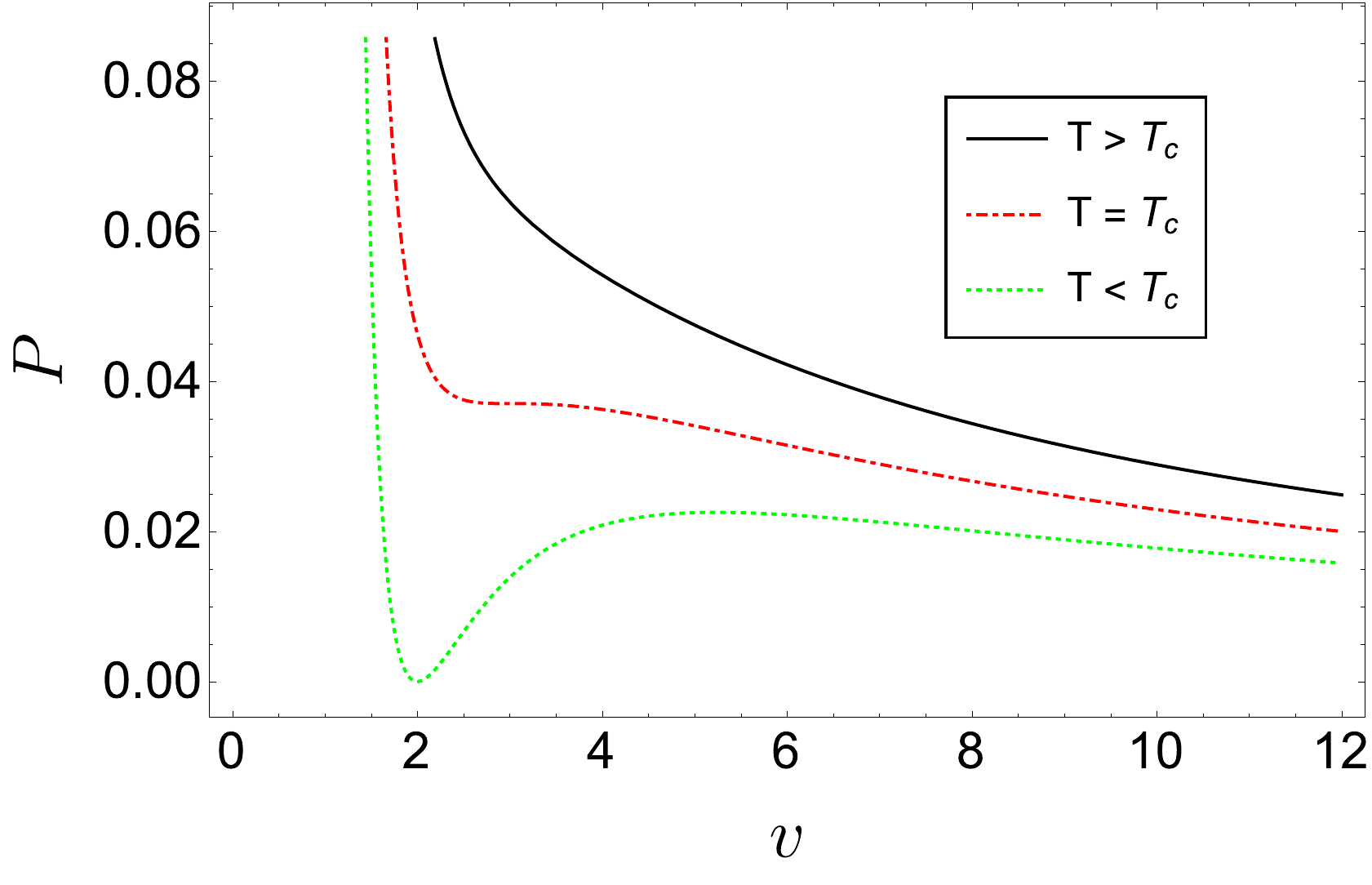}
\caption{$P-v$ diagram of van der Waals fluids. The temperature of isotherms decreases from top to bottom. The red dot-dashed line indicates the critical isotherm at $T=T_c$. We have set $a=b=1$ (online colors).}
\label{Fig0}
\end{center}
\end{figure}

Isotherms in the $P-v$ diagrams are shown in Fig.~\ref{Fig0}. Here we are only considering isotherms with $T>T_0=a/4b$. Below this threshold, which is obtained by imposing that the equation $P(v)=0$ has two distinct real roots, one can see that a branch of the isotherm has negative (i.e. unphysical) pressure~\cite{Kubiznak}. The inflection (critical) point of isotherms can be found by imposing the
following conditions
\begin{eqnarray}
\label{C1}
\left(\frac{\partial P}{\partial v}\right)_T&=&0\,,\\[2mm]
\left(\frac{\partial^2 P}{\partial v^2}\right)_T&=&0\,,
\label{C2}
\end{eqnarray}
which give
\begin{equation}
\label{cVan}
v_{c}=3b\,, \quad \,
T_{c}=\frac{8a}{27b}\,, \quad \,
P_{c}=\frac{a}{27b^2}\,.
\end{equation}
It is easy to verify that $P_c v_c/T_c=3/8$, which holds 
for all fluids, independently of the constants $a$ and $b$. 

The critical temperature $T_c$ marks a watershed. Indeed, for
$T<T_c$ the system undergoes a liquid-gas phase transition. To get
more knowledge about this phenomenon, we consider the Gibbs free
energy $G = G(P, T)$, which is given by integration of 
\be
\label{Gibbsdif} dG=-S dT + v dP\,, \ee for a fixed number of
particles. The specific Gibbs free energy is
\be
G(T,P)=-T\left\{1+\log\left[\frac{\left(v-b\right)T^{\frac{3}{2}}}{\lambda}\right]\right\}-\frac{a}{v}+P
v\,, 
\ee 
where $v$ is to be considered as a function of pressure
and temperature through van der Waals equation~\eqref{vdwequation}
and  $\lambda$ is a (dimensional) 
specific constant of the gas. 

\begin{figure}[t]
\begin{center}
\hspace{-3mm}\includegraphics[width=8cm]{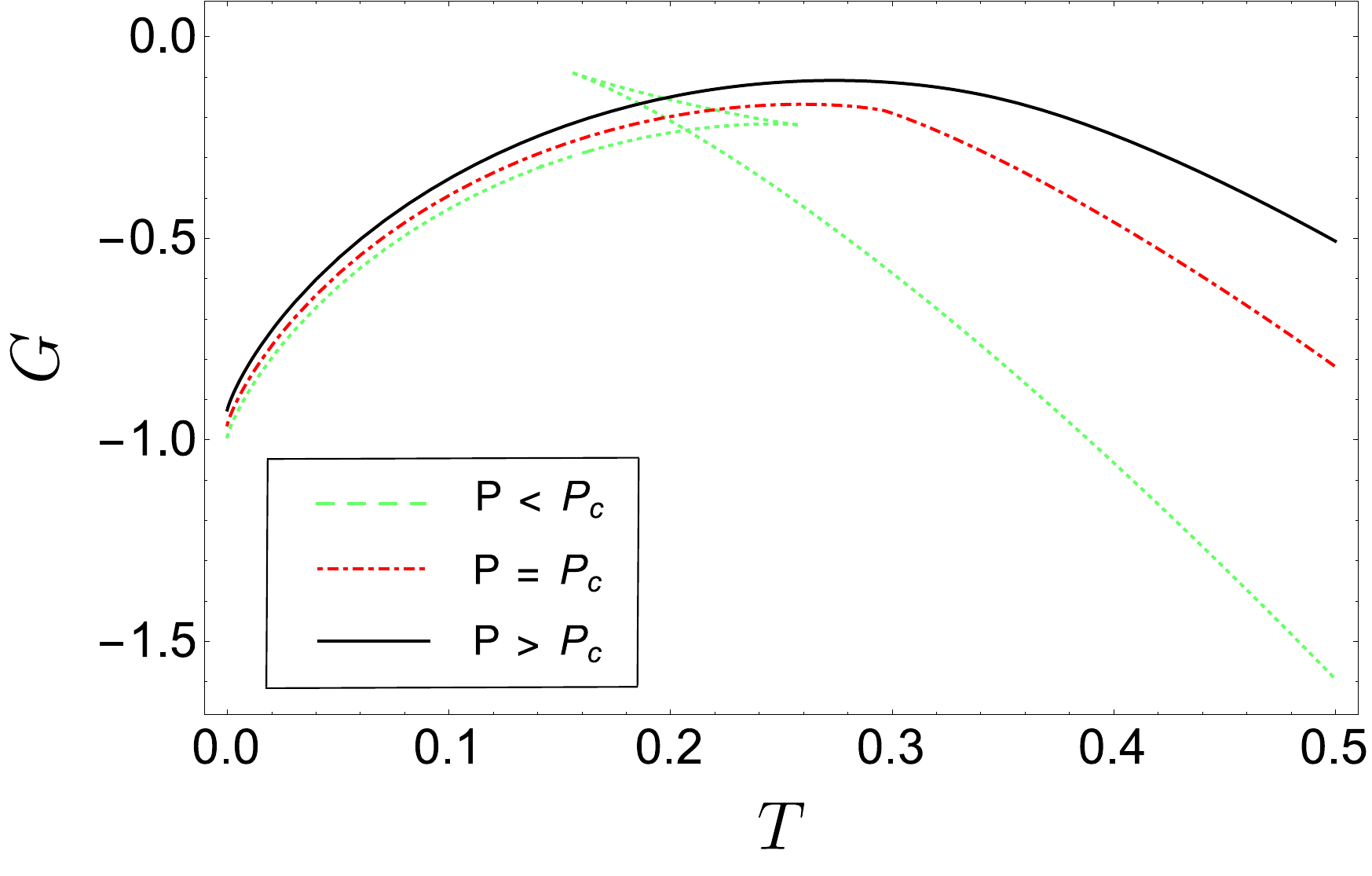}
\caption{Gibbs free energy as a function of $T$ for various $P$.
The red dot-dashed line indicates the critical isobar at $P=P_c$ (online colors).}
\label{Fig0bis}
\end{center}
\end{figure}

\begin{figure}[t]
\begin{center}
\hspace{-3mm}\includegraphics[width=8cm]{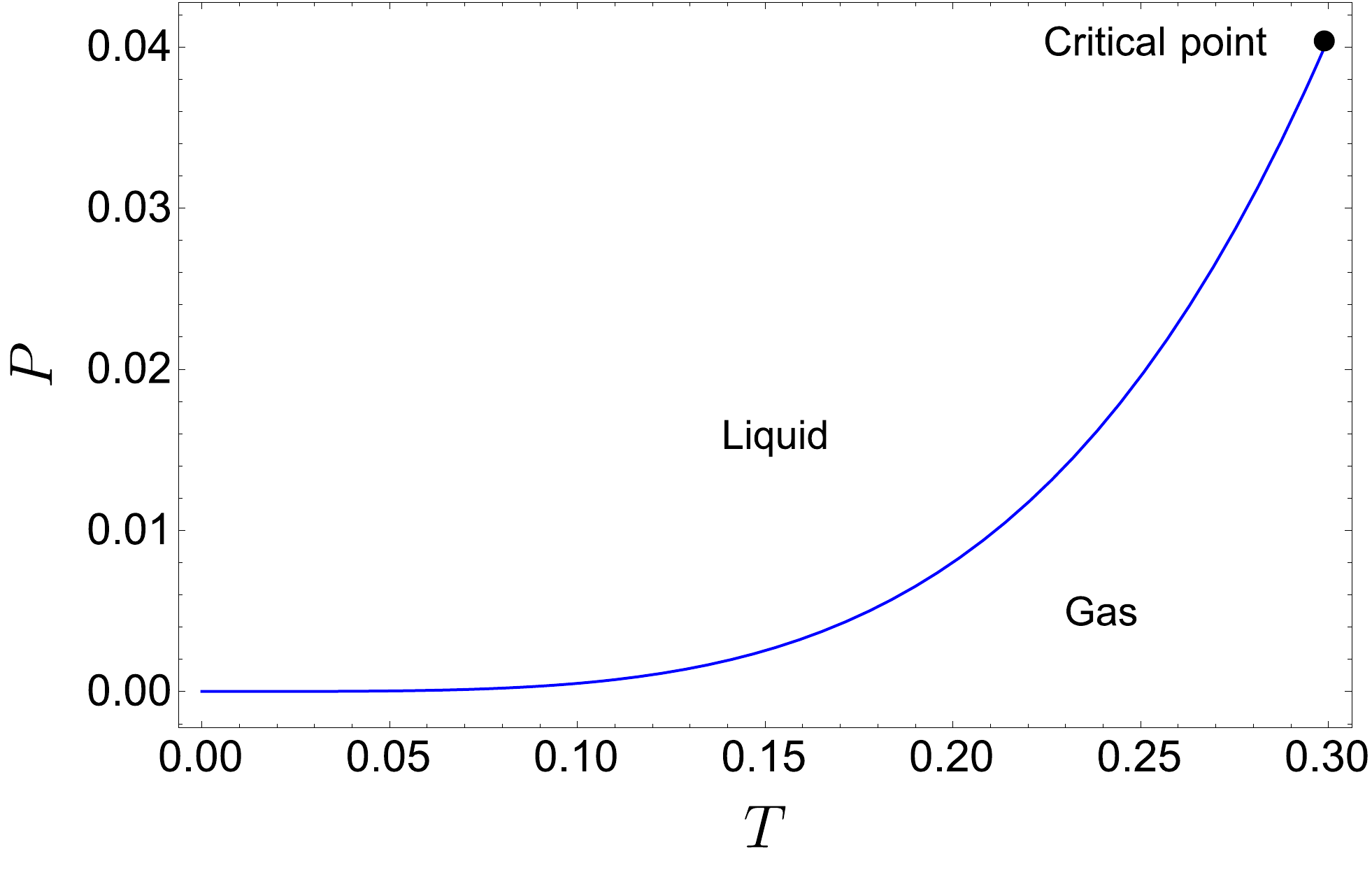}
\caption{Coexistence line of liquid-gas phase of van der Waals fluid in $P-T$ plane. The critical point is marked by a black circle.}
\label{Fig0ter}
\end{center}
\end{figure}

The Gibbs energy versus $T$ is shown in Fig.~\ref{Fig0bis} for different values of $P$. It can be seen that, below the critical pressure $P_c$ (green dotted curve), 
$G$ has a swallow tail behavior typical of first order
transitions from liquid to gas phases. The coexistence line along
which these two phases are in equilibrium occurs where there is a
crossing between two surfaces of $G$. This line is shown in
Fig.~\ref{Fig0ter} and obeys the Clausius-Clapeyron equation \be
\frac{dP}{dT}\bigg|_{coexist.}=\frac{S_g-S_l}{v_g-v_l}\,, \ee
where $S_{g,l}$ and $v_{g,l}$ denote the specific entropy and
volume of the gas and liquid phases, respectively. This line
terminates at the critical point, beyond which it is no longer
possible to distinguish the two phases. 

Let us also comment on the characteristic critical exponents,
which feature the behavior of the system near the
critical point.  For this purpose, we introduce the variables
\be
\label{TildeQ} \tilde T=\frac{T-T_c}{T_c}\,, \qquad \tilde
v=\frac{v-v_c}{v_c}\,. 
\ee 
The basic critical exponents
$\alpha,\beta,\gamma$ and $\Delta$ are defined as follows:
\begin{itemize}
\item [-] $\alpha$ rules the dynamics of the specific heat at constant volume $C_v$ according to
\be
\label{alpha}
C_v=T\left(\frac{\partial S}{\partial T}\right)_v \propto |\tilde T|^{-\alpha}\,.
\ee
\item[-] $\beta$ governs the behavior of the order parameter $\eta=v_g-v_l$
for each isotherm as
\be
\label{beta}
\eta\propto |\tilde T|^{\beta}\,.
\ee
\item[-] $\gamma$ is associated to the isothermal compressibility $\kappa_T$ according to
\be
\label{gamma}
\kappa_T=-\frac{1}{v}\left(\frac{\partial v}{\partial P}\right)_T\propto |\tilde T|^{-\gamma}\,.
\ee
\item[-] $\Delta$ fixes the behavior of the difference $|P-P_c|$ on the critical isotherm $T=T_c$ as
\be
\label{Delta}
|P-P_c|\propto|v-v_c|^{\Delta}\,.
\ee
\end{itemize}

Details of calculations can be found in~\cite{Gold} (see also the
related discussion for BHs in Sec.~\ref{critpar}). Here, we 
exhibit final results and briefly comment on their
derivation. By explicit computation of $C_v$, one can show
that it does not depend on $\tilde T$. In turn, this implies
$\alpha=0$ from Eq.~\eqref{alpha}. 
On the other hand, considerations on the equation of
corresponding states and Maxwell's equal area law lead to
$\beta=1/2$ and $\gamma=1$, while the study of the
shape of the critical isotherm yields $\Delta=3$.

In the next Section we repropose the above study in the context of Tsallis entropy-based thermodynamics of BHs. We discuss how BH critical phenomena are affected by the prescription~\eqref{TsEn}
both at qualitative and quantitative levels, highlighting
similarities and differences with phase transitions of van der Waals fluids.

\section{Phase transitions of charged AdS black holes with global monopole}
\label{Pht}

Let us start by reviewing the thermodynamic
properties of charged AdS BHs with global monopole.
In~\cite{Barriola:1989hx} Barriola and Vilenkin have shown
that the simplest lagrangian model that accounts for the creation
of global monopole is
\be
\label{lag}
\mathcal{L}_{gm}=\frac{1}{2}\partial_{\mu}\Phi^a\partial^{\mu}\Phi^{*a}
-\frac{\gamma}{4}\left(\Phi^a\Phi^{*a}-\eta_0^2\right)^2\,,
\ee
where $\Phi^a$ is a triplet of scalar fields, $\gamma$ the coupling
constant (not to be confused with the critical exponent in Eq.~\eqref{gamma}) and $\eta_0$ the parameter associated to the
symmetry breaking energy scale\footnote{Following the approach of~\cite{Rani:2022xza,Jawad:2022lww}, here we suppose that Tsallis  prescription only influences BH entropy, while leaving the lagrangian and field equations of the theory unaffected. A more in-depth study of this technical aspect shall be reserved for the future.}. The field configuration 
of the scalar field triplet is described by
\be
\Phi^a=\eta_0h(\tilde r)\frac{\tilde x^a}{\tilde r}\,,
\ee
where $\tilde x^a\tilde x^a=\tilde r^2$\,.

The general static spherically symmetric metric that describes a four-dimensional AdS BH with global monopole is
\be
\label{metric}
d\tilde s^2=-\tilde f(\tilde r)d\tilde t^2+\tilde f(\tilde r)^{-1}d\tilde r^2+\tilde r^2d\Omega^2\,,\quad\, 
\ee
where $d\Omega^2=d\theta^2+\sin^2\theta d\phi^2$ is the angular part
of the metric on the two sphere. The field equation for the scalar
field $\Phi^a$ in the above metric reads
\be
\tilde f h''+2\tilde f\frac{h'}{\tilde r}+\tilde f' h'-2\frac{h}{\tilde r^2}-\gamma\eta_0^2h(h^2-1)=0\,,
\ee
where the prime denotes derivative with respect to $\tilde r$.

In certain approximation, one can show that the solution for a charged AdS BH with global monopole and the gauge potential takes the form~\cite{Soroushfar:2020wch}
\begin{eqnarray}
\tilde f(\tilde r)&=&1-8\pi\eta_0^2-\frac{2\tilde m}{\tilde r}+\frac{\tilde q^2}{\tilde r^2}+\frac{\tilde r^2}{l^2}\,,\\[2mm]
&&\hspace{7.5mm}\tilde A=\frac{\tilde q}{\tilde r}d\tilde t\,,
\end{eqnarray}
where $\tilde m$ and  $\tilde q$ are the mass and electric charge
parameters, respectively, while $l$ is the AdS radius related to
the cosmological constant by $\Lambda=-3/l^2$.

It is now convenient to introduce the following coordinate rescaling
\begin{eqnarray}
\tilde t&=&\left(1-8\pi\eta_0^2\right)^{-\frac{1}{2}}t\,,\\[2mm] 
\tilde r&=&\left(1-8\pi\eta_0^2\right)^{\frac{1}{2}}r\,,
\end{eqnarray}
along with the definitions
\begin{eqnarray}
m&=&\left(1-8\pi\eta_0^2\right)^{-\frac{3}{2}}\tilde m\,,\\[2mm]
q&=&\left(1-8\pi\eta_0^2\right)^{-1}\tilde q\,,\\[2mm]
\eta^2&=&8\pi\eta_0^2\,.
\end{eqnarray}

In terms of these new quantities, the line element~\eqref{metric} and the gauge potential become
\begin{eqnarray}
ds^2&=&-f(r)d t^2+ f( r)^{-1}d r^2+\left(1-\eta^2\right) r^2d\Omega^2\,,\\[2mm]
&&\hspace{18mm}A=\frac{q}{r}dt\,,
\end{eqnarray}
where
\be
f(r)=1-\frac{2 m}{ r}+\frac{ q^2}{ r^2}+\frac{ r^2}{l^2}\,.
\ee
Also, the electric charge and Arnowitt-Deser-Misner (ADM) mass read
\begin{eqnarray}
Q&=&\left(1-\eta^2\right)q\,,\\[2mm]
M&=&\left(1-\eta^2\right)m\,.
\end{eqnarray}
From the latter equation, we notice that $\eta^2$ must be less than unity in order for $M$ to be positively defined. 

To introduce the thermodynamic variables of AdS BHs at equilibrium,
let us observe that the BH event horizon located at $r_h$
is given by the largest root of $f(r)=0$. In turn, this condition
can be exploited to express the ADM mass as
\be
\label{mass}
M(r_h)=\frac{(1-\eta^2)}{2}r_h + \frac{Q^2}{2r_h(1-\eta^2)}+\frac{r^3_h(1-\eta^2)}{2l^2}\,.
\ee

The surface gravity can be inferred from the relation
$\kappa^2=-\xi_{\mu;\nu}\xi^{\mu;\nu}/2|_{r=r_h}$, where
$\xi^{\mu}$ is the timelike Killing vector $(\partial_t)^\mu$.
This gives~\cite{Poisson}
\be
\kappa(r_h)=\frac{1}{2r_h}\left[1+\frac{3r_h^2}{l^2}-\frac{Q^2}{(1-\eta^2)^2r_h^2}\right].
\ee
Accordingly, Hawking temperature is defined by $T=\kappa/2\pi$.
Furthermore, the surface area $A_{bh}$ of the BH horizon  is
\be
A_{bh}=\int_{r=r_h}\sqrt{g_{\theta\theta}\,g_{\phi\phi}}d\theta d\phi =4\pi\left(1-\eta^2\right)r_h^2\,.
\ee

Now, within the standard framework of Boltzmann-Gibbs statistics,
BH entropy obeys the well-known area law
\be
S_{BH}=\frac{A_{bh}}{4}=\pi\left(1-\eta^2\right)r_h^2\,.
\ee
However, following non-additive Tsallis recipe~\eqref{TsEn},
this relation should be generalized  to~\cite{TsallisCirto}
\be
\label{TsaEN}
S_{BH}\rightarrow S_\delta= \left(S_{BH}\right)^\delta =\left[\pi\left(1-\eta^2\right)r_h^2\right]^{\delta}\,,
\ee
which introduces the non-additive 
$\delta$-degree of freedom in our model\footnote{Since departure from Boltzmann-Gibbs statistics is expected to be relatively small, we focus on deviations $|\delta-1|< 1/2$. This assumption is supported by many recent constraints in literature~\cite{Luciano:2022ely}. Furthermore, we shall see later that it is corroborated by physical requirements on BH critical parameters.}.
To streamline the notation, henceforth we shall simply denote
Tsallis entropy by $S$.  Notice also that
a similar analysis for the case of five-dimensional Schwarzschild AdS BHs in $AdS_5\times S^5$ spacetime has been recently proposed in~\cite{Rani:2022xza} neglecting global monopole effects.

In the extended phase space, the thermodynamic pressure and volume of asymptotically AdS BHs are identified with the cosmological constant and its conjugate quantity~\cite{Kastor:2009wy}, i.e.
\begin{eqnarray}
\label{pressure}
P&=&-\frac{\Lambda}{8\pi}=\frac{3}{8\pi l^2}\,,\\[2mm]
V&=&\left(\frac{\partial M}{\partial P}\right)_{S,Q}
=\frac{4}{3}\pi\left(1-\eta^2\right)r_h^3\,.
\label{Vol}
\end{eqnarray}
These identifications allow to formulate a thermodynamics
of BHs in the extended phase space, leading to the
emergence of phase transitions with low-energy counterparts
in van der Waals fluids~\cite{Chamblin:1999tk,Chamblinbis}, provided one works in the canonical (fixed charge) ensemble.

The above considerations indicate that, in the presence of a global monopole in non-additive entropy regime, the new thermodynamic variables are non-trivially affected by the symmetry breaking and 
Tsallis parameters.
Clearly, the standard definitions for four-dimensional Reissner-Nordstr\text{\"o}m AdS BHs are recovered
for $\eta=0$ and $\delta=1$.
In spite of the modified setting, it is easy to verify that Eqs.~\eqref{mass}-\eqref{Vol} still obey the first law of BH thermodynamics
\be
\label{FLT}
dM=TdS +\varphi dQ+VdP\,,
\ee
and Smarr relation
\be
\label{Smarr}
M = 2\left(TS-VP\right)+\varphi Q\,,
\ee
where $T=\left(\frac{\partial M}{\partial S}\right)_{P,Q}$ and $\varphi=\left(\frac{\partial M}{\partial Q}\right)_{S,P}$ are the temperature and electric potential, respectively. 

\begin{figure}[t]
\begin{center}
\hspace{-3mm}\includegraphics[width=8cm]{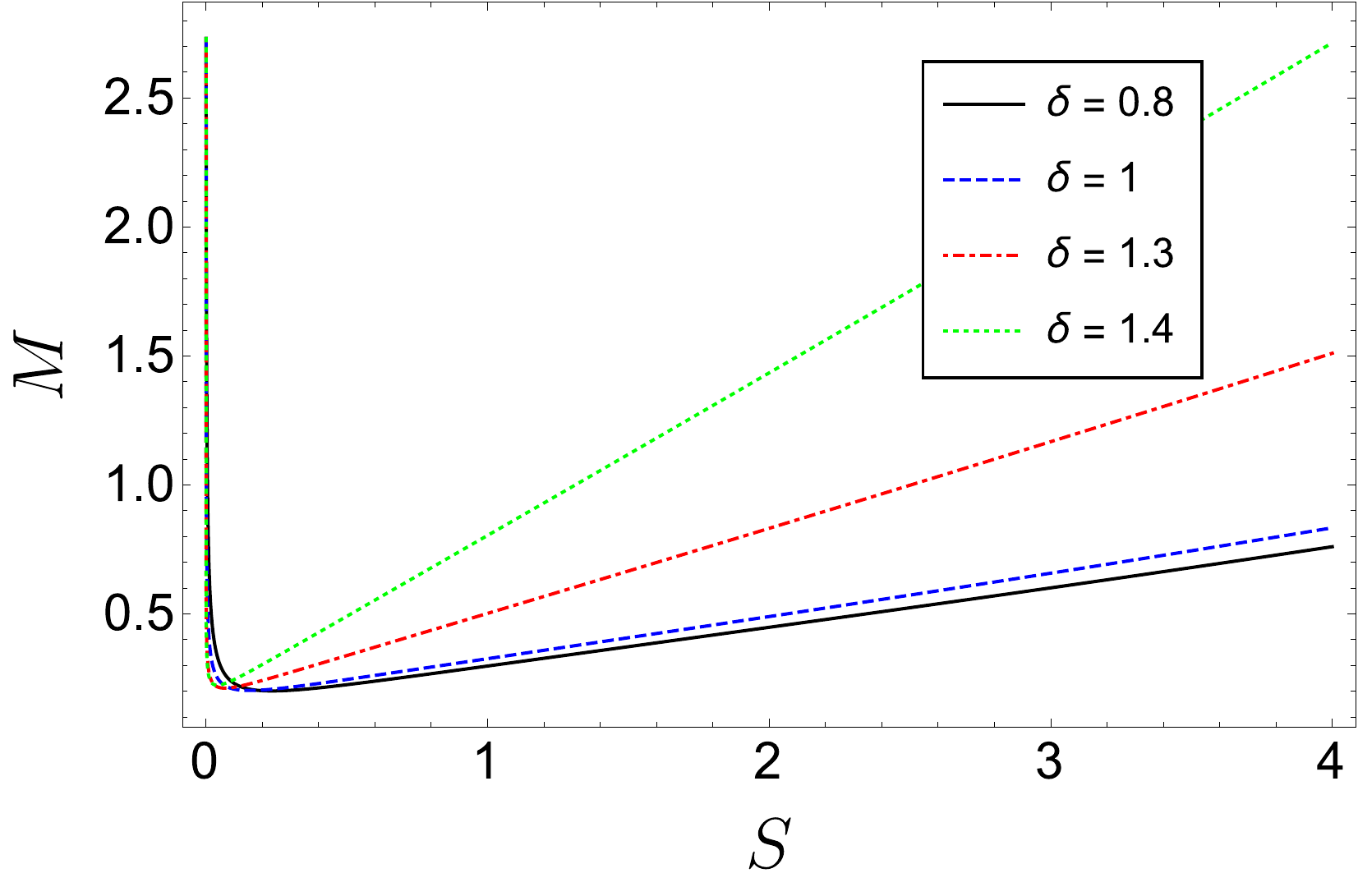}
\\[0.5cm]
\hspace{-3mm}\includegraphics[width=8cm]{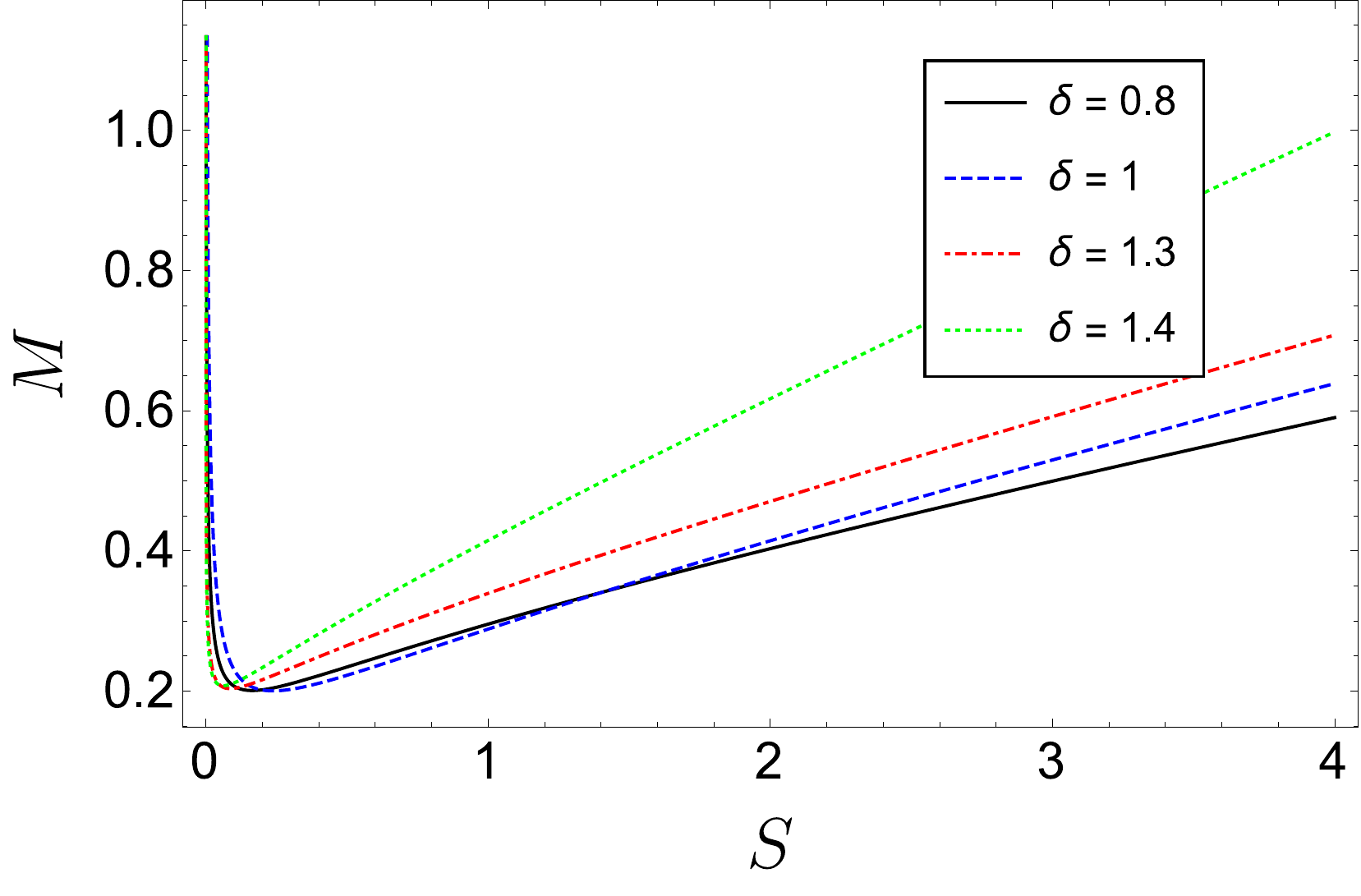}
\\[0.5cm]
\hspace{-3mm}\includegraphics[width=8cm]{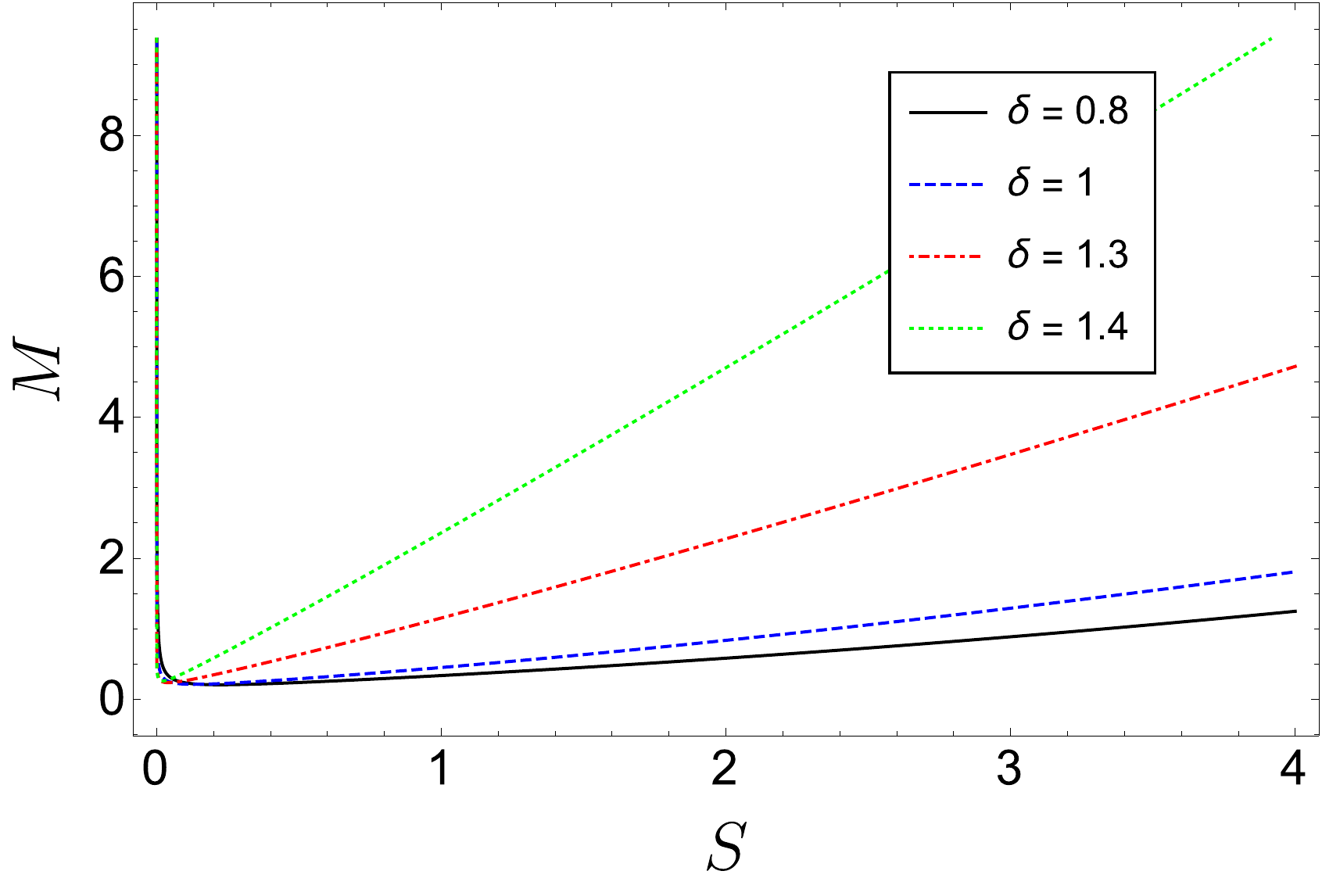}
\caption{Plot of $M$ as a function of $S$ for various values of $\delta$ and $l=l_c$ fixed through the critical condition~\eqref{Pc} (upper panel), $l=2l_c$ (middle panel) and $l=0.5l_c$ (lower panel) (online colors).}
\label{Fig1}
\end{center}
\end{figure}

Equations~\eqref{FLT} and~\eqref{Smarr} along with the
deformed entropy~\eqref{TsaEN} provide the basic
tools for our next analysis. According to our previous discussion, 
we shall work in the canonical ensemble, treating the BH charge $Q$ as a fixed external parameter, rather than a thermodynamic variable.
Following, for example, the approach of~\cite{Rani:2022xza,Soroushfar:2020wch,Jawad:2022lww}, we rearrange the ADM mass parameter~\eqref{mass} in terms of non-additive entropy~\eqref{TsaEN} as
\be
\label{MassAdSBH}
M(S)=\frac{S^{-\frac{1}{2\delta}}\left\{S^{\frac{2}{\delta}}+\pi l^2 \left[\pi Q^2+S^{\frac{1}{\delta}}\left(1-\eta^2\right)\right]\right\}}
{2l^2\left[\pi^3\left(1-\eta^2\right)\right]^{\frac{1}{2}}}\,.
\ee

The behavior of $M$ versus $S$ is plotted in Fig.~\ref{Fig1} for various values of $\delta,l$ and fixed $Q=0.2, \eta=0.5$ as in~\cite{Deng:2018wrd,Jawad:2022lww} (unless specified otherwise, we shall keep these values fixed throughout the whole analysis).
We can see that $M$ initially decreases, then attains a minimum value and finally starts increasing again. This implies that BH
evaporates emitting Hawking radiation up to a certain value of $S$
(i.e. $r_h$), after that net absorption prevails.
Effects of Tsallis entropy manifest through a variation of the growth rate of $M$. In particular, for $\delta<1$ (concave Tsallis entropy), the mass increases slower than the standard $\delta=1$ case, while the reverse happens for $\delta>1$.

In Hawking thermodynamic picture,
BHs are expected to emit thermal radiation in a continuous spectrum according to their temperature~\cite{Hawking:1975vcx}. From the first law of thermodynamics, we have
\begin{eqnarray}
\nonumber
\hspace{-3mm}T(S)&=&\left(\frac{\partial M}{\partial S}\right)_{P}\\[2mm]
&=&\frac{S^{-\frac{2\delta+1}{2\delta}}
\left\{3S^{\frac{2}{\delta}}-\pi l^2\left[\pi Q^2-S^{\frac{1}{\delta}}\left(1-\eta^2\right)\right]
\right\}
}
{4\,\delta\,l^2\left[\pi^3\left(1-\eta^2\right)\right]^{\frac{1}{2}}}\,.
\label{TempS}
\end{eqnarray}

\begin{figure}[t]
\begin{center}
\includegraphics[width=8.1cm]{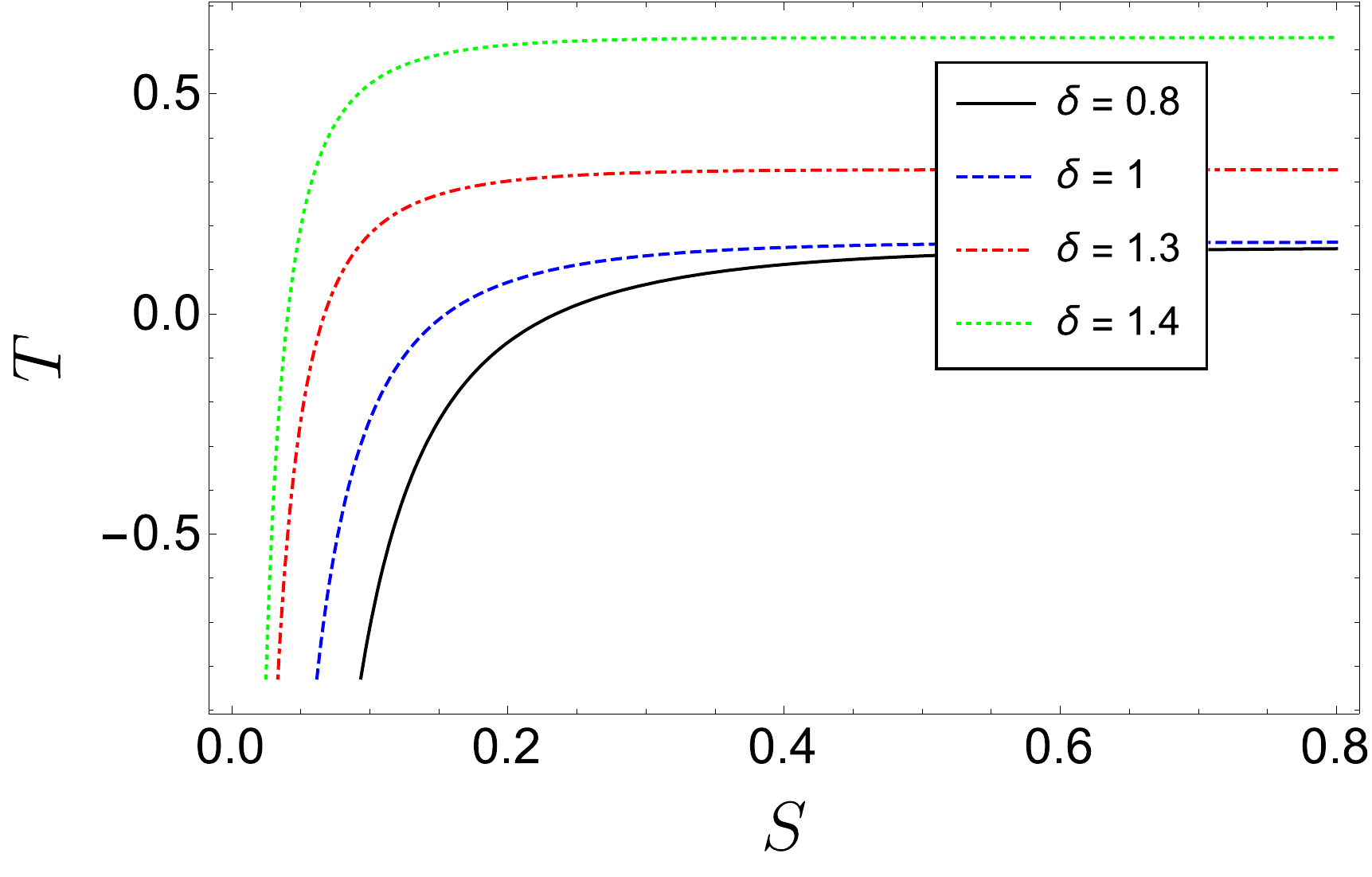}
\\[0.5cm]
\includegraphics[width=8.2cm]{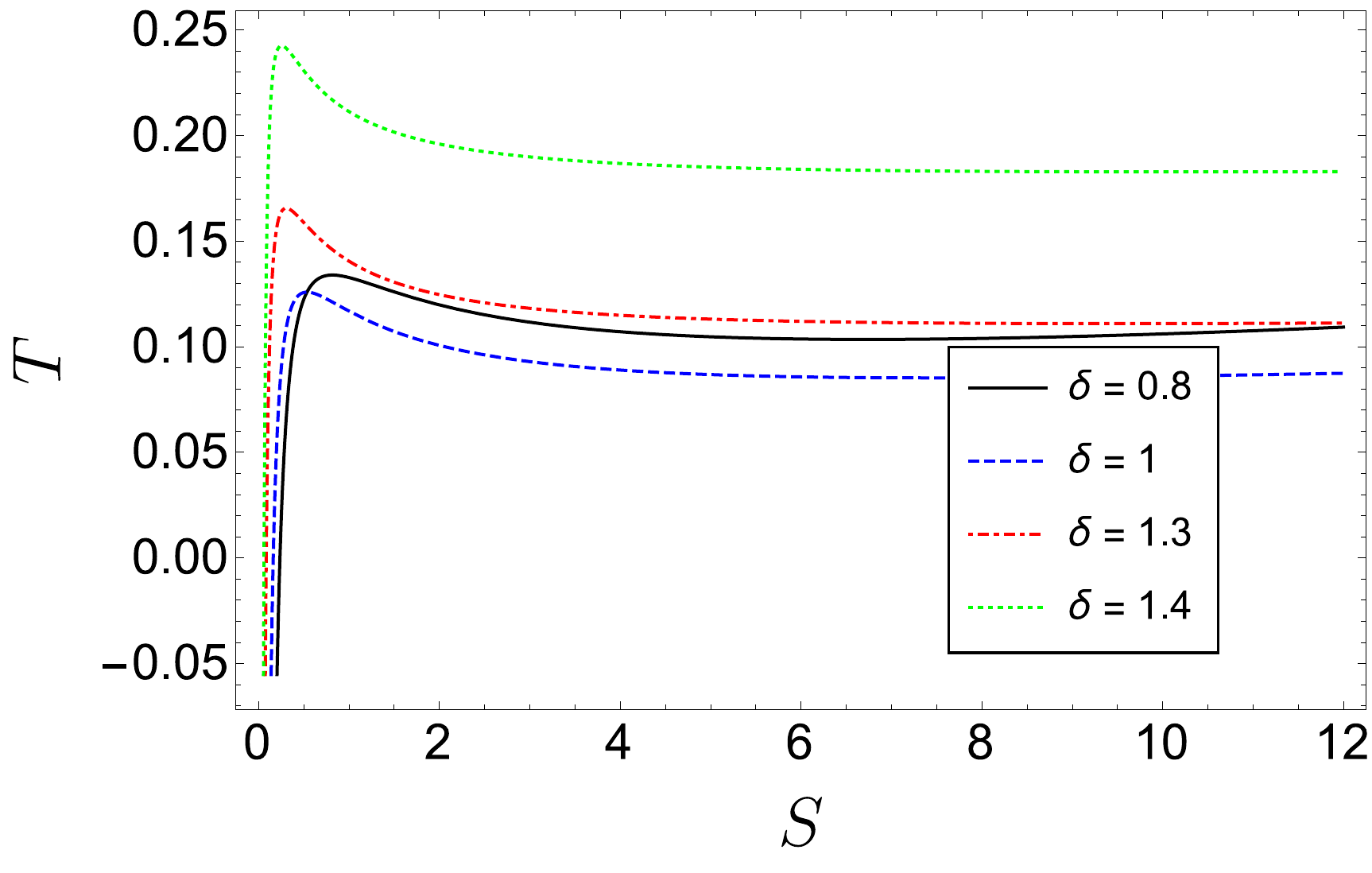}
\\[0.5cm]
\hspace{2mm}\includegraphics[width=8cm]{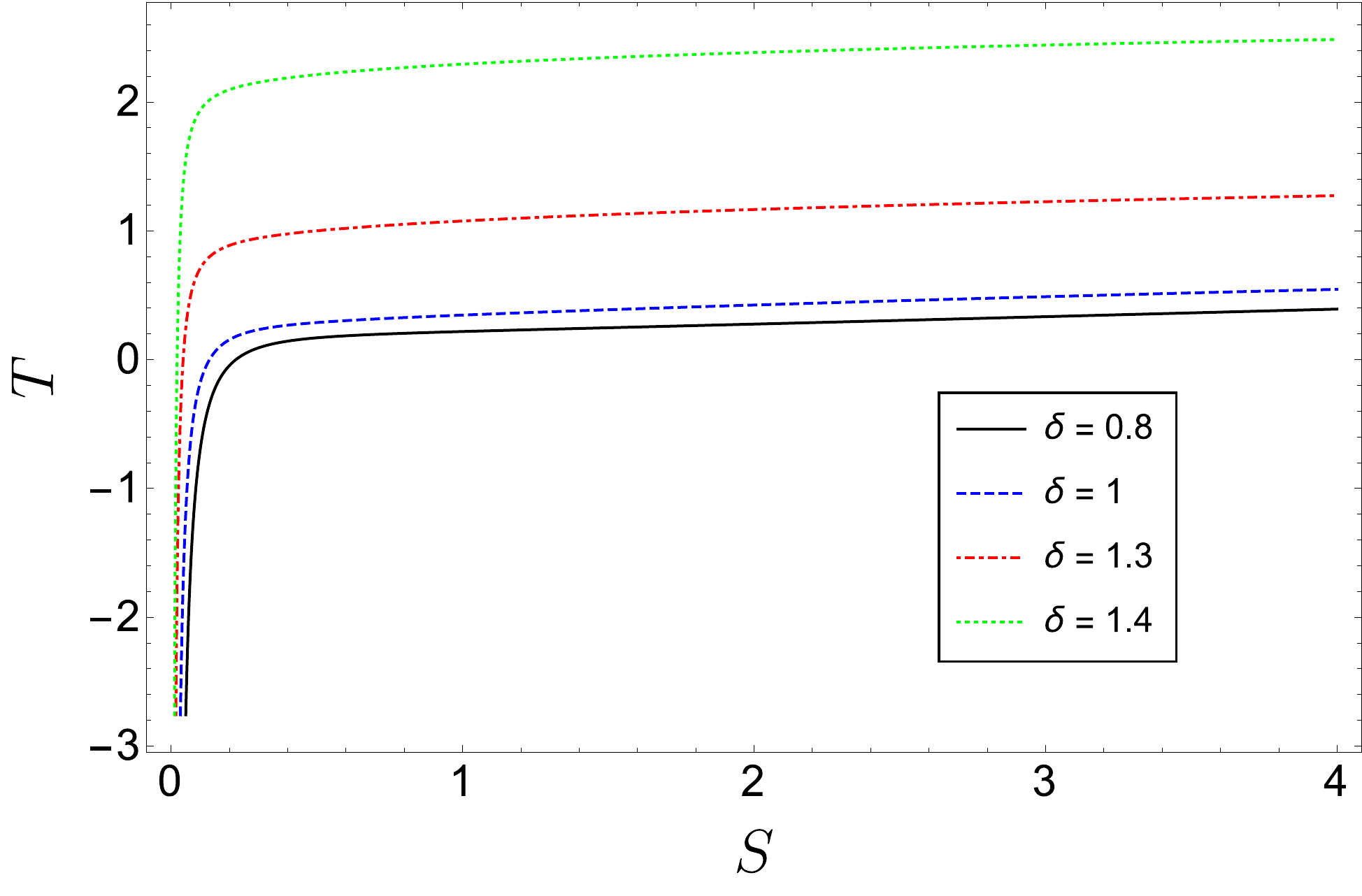}
\caption{Plot of $T$ as a function of $S$ for various values of $\delta$ and $l=l_c$ fixed through the critical condition~\eqref{Pc} (upper panel), $l=2l_c$ (middle panel) and $l=0.5l_c$ (lower panel) (online colors).}
\label{Fig2}
\end{center}
\end{figure}

Figure~\ref{Fig2} displays the behavior of $T$ versus $S$
for various values of $\delta,l$ and fixed $Q,\eta$. Since the slope of the $T-S$ graph is related to the specific heat (see the next section for more quantitative discussion), it provides us with information on the stability of BHs with respect to fluctuations. We can see that $T$ increases monotonically, which denotes stable behavior, and has no (lower panel) or a unique (upper panel) stationary point, depending on the value of $l$. The entropy $S_*$ corresponding to this stationary point is obtained as
\begin{eqnarray}
\label{Sstar}
\hspace{-3mm}\left(\frac{\partial T}{\partial S}\right)_{P}(S_*)&=&0\,\, \Longrightarrow\,\,S_*=
\left[\frac{2\pi\hspace{0.2mm}Q^2\left(1+2\delta\right)
}
{(2\delta-1)\left(1-\eta^2\right)}
\right]^{\delta}\,,
\end{eqnarray}
which is clearly $\delta$-dependent.
Should we retrace accurately the $T-S$ (or $T-r_h$) profile of BHs, we could establish to what extent BH thermodynamics
obeys Tsallis prescription and possibly constrain the parameter $\delta$.
On the other hand, from the middle panel of Fig.~\ref{Fig2}, 
we see that $T$ increases for small and large entropies, while
exhibiting a decreasing (unstable) behavior in the intermediate domain. In this case, two stationary points occur, which are given by
\begin{eqnarray}
\label{dobst}
S_{*1,2}&=&
\left[\frac{\pi l^2\left(1-2\delta\right)\left(1-\eta^2\right)\pm \pi {g_{l,Q,\eta,\delta}^{\frac{1}{2}}}}
{6(2\delta-3)}
\right]^{\delta}.
\end{eqnarray}
Here we have defined 
\begin{eqnarray}
\nonumber
\hspace{-4mm}g_{l,Q,\eta,\delta}&\equiv&\\[2mm]
&&\hspace{-17mm} l^2\left[12Q^2\left(2\delta-3\right)\left(1+2\delta\right)+\,l^2\left(1-2\delta\right)^2\left(1-\eta^2\right)^2
\right].
\label{f}
\end{eqnarray}

In the next section, we show that these 
stationary points signal a critical behavior of BHs, which can be described using the language of liquid-gas phase transitions of van der Waals fluids. 
Moreover, we observe that in all the plots in Fig.~\ref{Fig2}, the root 
\begin{eqnarray}
\label{plp}
T(S_0)&=&0\,
\Longrightarrow\,\\[2mm]
\nonumber
&&\hspace{-15mm}S_0=\left\{\frac{\pi}{6}\left[\left[12l^2Q^2+l^4\left(1-\eta^2\right)^2\right]^{\frac{1}{2}}-l^2\left(1-\eta^2\right)\right]\right\}^\delta\,,
\end{eqnarray}
depicts a boundary line between non-physical ($T<0$) and physical $(T>0)$ domains, which is usually referred to as physical limitation point. The entropy corresponding to the physical limitation temperature can be used to infer the value of the limiting horizon radius via Eq.~\eqref{TsaEN}. Also, we have $T(S\rightarrow\infty)\rightarrow\infty$,  regardless of the values of $\delta$ and $l$,
which indicates stability of BHs in this regime.

\subsection{Critical points of phase transitions}
\label{CritP}

To study critical phenomena and phase transitions of charged AdS BHs, we move to investigate another relevant thermodynamic quantity, namely the specific heat at constant pressure 
\be
\label{defCP}
C_p=T\left(\frac{\partial S}{\partial T}\right)_{P}\,.
\ee
In general, a positive specific heat allows a stable BH to exist, while negative values indicate that BHs will disappear when suffering from a small perturbation. Moreover, singularities of $C_p$ potentially correspond to phase transition critical points.

\begin{figure}[t]
\begin{center}
\hspace{-6mm}\includegraphics[width=8.7cm]{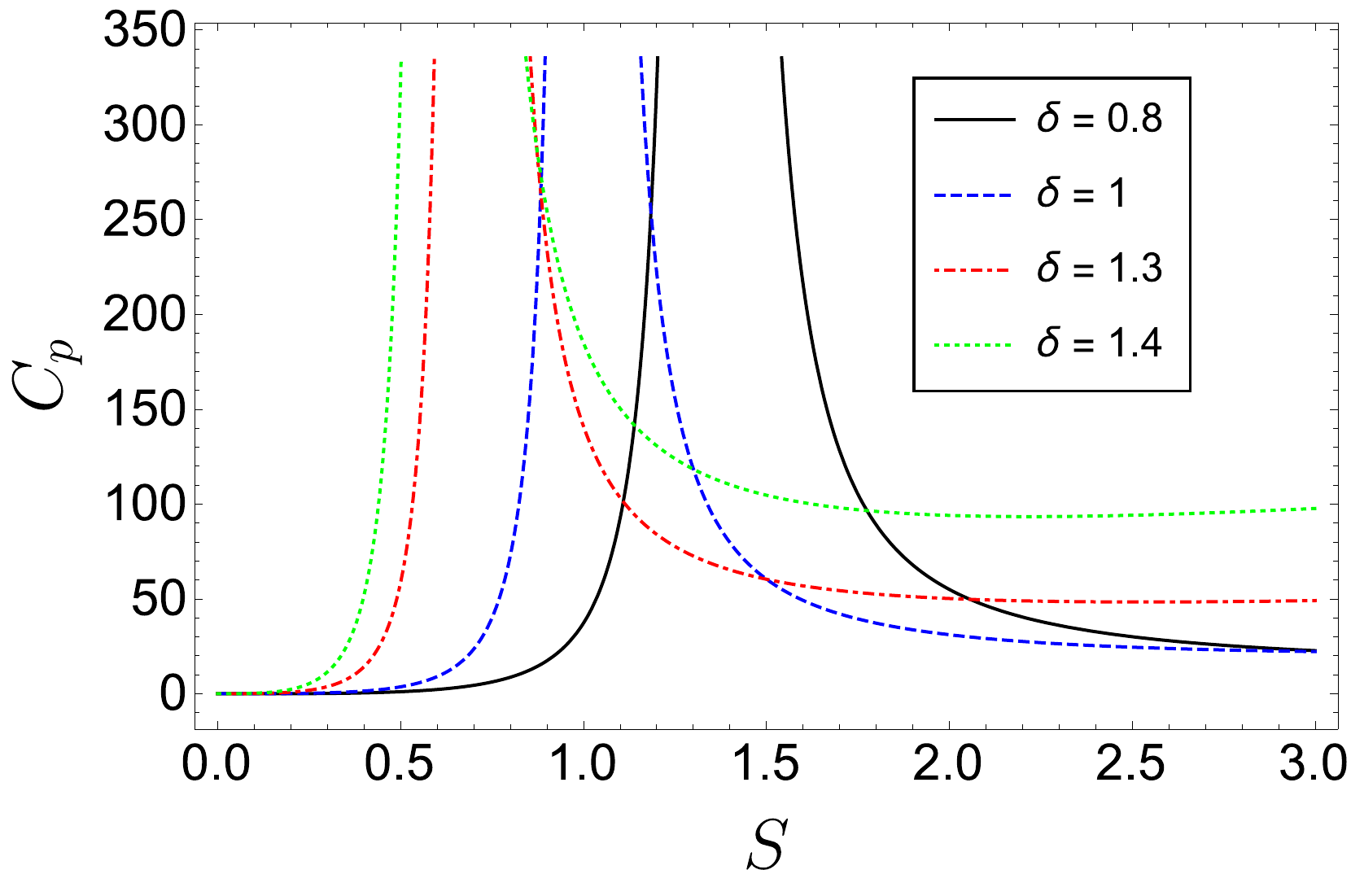}
\\[0.5cm]
\hspace{-6mm}\includegraphics[width=8.4cm]{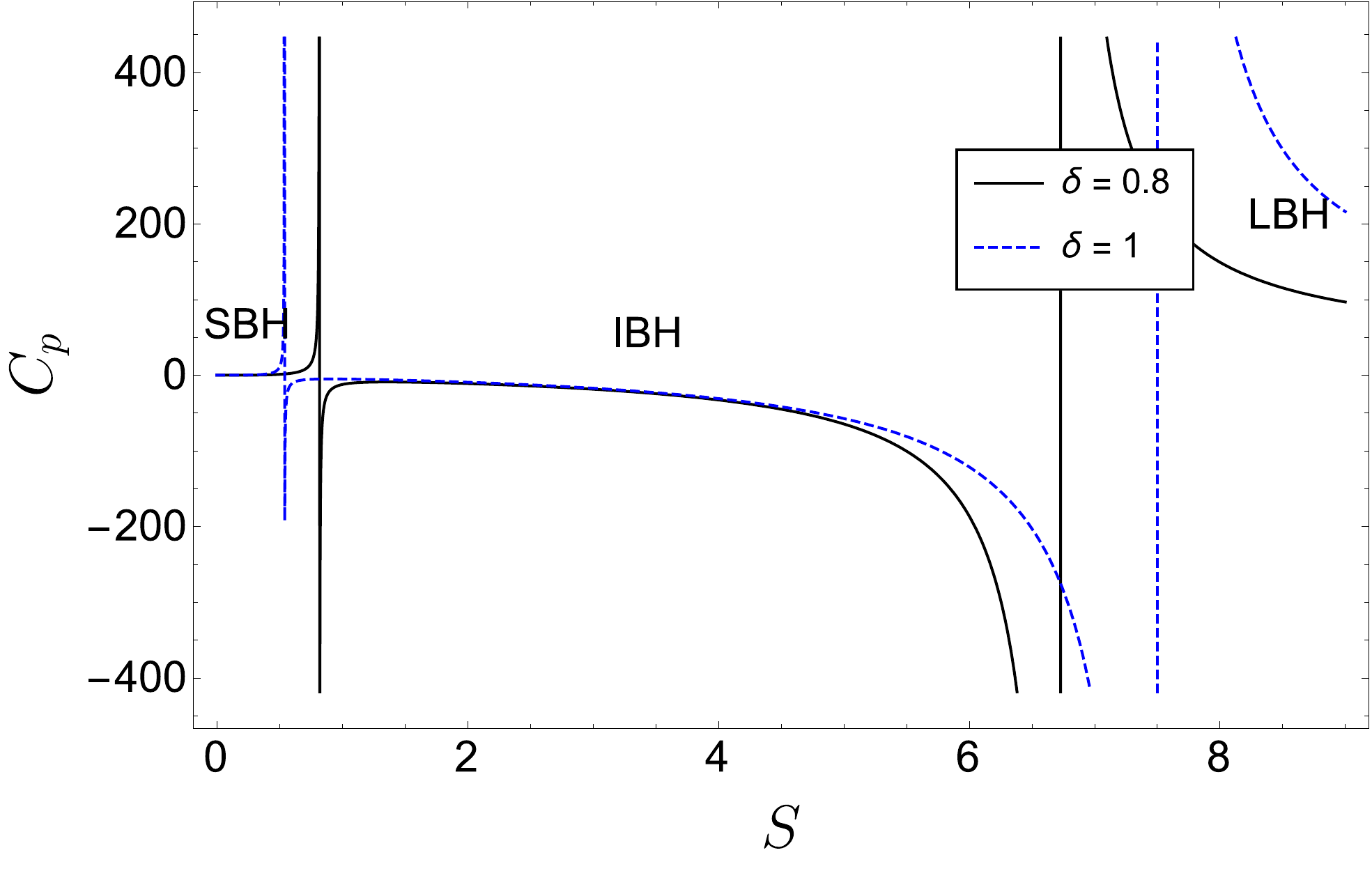}
\\[0.5cm]
\hspace{-5mm}\includegraphics[width=8.3cm]{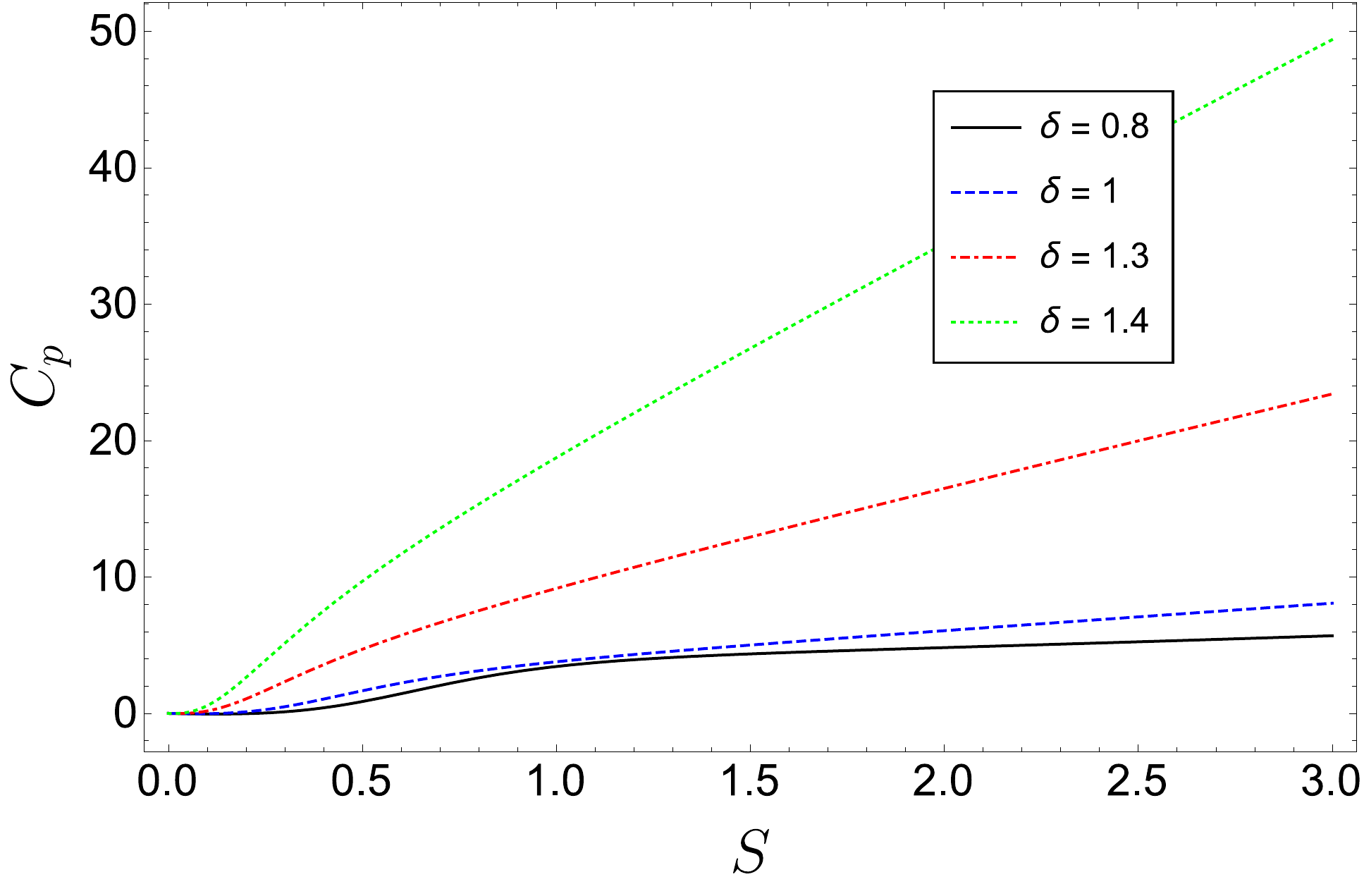}
\caption{Plot of $C_p$ as a function of $S$ for various values of $\delta$ and $l=l_c$ fixed through the critical condition~\eqref{Pc} (upper panel), $l=2l_c$ (middle panel) and $l=0.5l_c$ (lower panel). For visual clarity, we have only considered $\delta=0.8$ and $\delta=1$ in the middle panel (online colors).}
\label{Fig3}
\end{center}
\end{figure}

From Eq.~\eqref{TsaEN} and~\eqref{TempS}, we get
\begin{eqnarray}
\label{Cp}
C_p(S)&=&\\[2mm]
\nonumber
&&\hspace{-15mm}\frac{2\, \delta S\left\{\pi l^2\left[\pi Q^2-S^{\frac{1}{\delta}}\left(1-\eta^2
\right)
\right]-3S^{\frac{2}{\delta}}
\right\}}
{3S^{\frac{2}{\delta}}(2\delta-3)-\pi l^2\left[\pi Q^2(1+2\delta)-S^{\frac{1}{\delta}}(2\delta-1)(1-\eta^2)
\right]
}.
\end{eqnarray}
The behavior of $C_p$ as a function of $S$
is plotted in Fig.~\ref{Fig3} for various values of $\delta,l$ and fixed $Q,\eta$. Consistently with the discussion below Eq.~\eqref{TempS}, it is observed that $C_p$ exhibits one (upper panel), two (middle panel) or no (lower panel) discontinuity, depending on the value of $l$. These discontinuities are identified by vanishing slope of $T-S$ diagrams (see also Eqs.~\eqref{Sstar} and~\eqref{dobst}). 
From the middle panel, we clearly distinguish three regions.
Both the small and large
entropy (i.e. radius) regimes 
are thermodynamically stable ($C_p>0$), while
the intermediate one is unstable ($C_p<0$), implying
a transition between the small BH (SBH) and large BH (LBH). Apparently, Tsallis entropy affects the width of the entropy
interval where such a transition
occurs. Also, we observe that 
the intermediate region degenerates in a single point 
as $l$ increases above a certain critical value, independently of $\delta$ (upper panel). On the other hand, for $l$ below this threshold, $C_p$ does not exhibit any discontinuity and always remains positive (lower panel),  which means that BHs are locally stable and no transition takes place. 

To understand small-large BH phase transitions
at a more quantitative level, let us derive
a geometric-like equation of state for BHs in the extended phase space. Toward this end, we combine the definitions~\eqref{pressure} and~\eqref{TempS} of thermodynamic pressure and temperature to obtain
\be
\label{eqgeom}
P(r_h)=\frac{\delta\left[\pi\left(1-\eta^2\right)\right]^{\delta-1}T}{2\,r_h^{3-2\delta}} +\frac{Q^2}{8\pi\left(1-\eta^2\right)^2r_h^4}-\frac{1}{8\pi r_h^2}\,. 
\ee
Comparison with van der Waals equation of state allows
us to convert the above geometric relation into a physical one
by identifying the horizon radius $r_h$
with the specific volume of van der Waals fluid as~\cite{Kubiznak}
\be
\label{defvrh}
v=2 r_h\,.
\ee
This yields
\be
\label{eqphys}
P(v)=\frac{2^{2\left(1-\delta\right)}\delta\left[\pi\left(1-\eta^2\right)\right]^{\delta-1}T}{v^{3-2\delta}} +\frac{2Q^2}{\pi\left(1-\eta^2\right)^2v^4}-\frac{1}{2\pi v^2}\,.
\ee

\begin{figure}[t]
\begin{center}
\includegraphics[width=7.6cm]{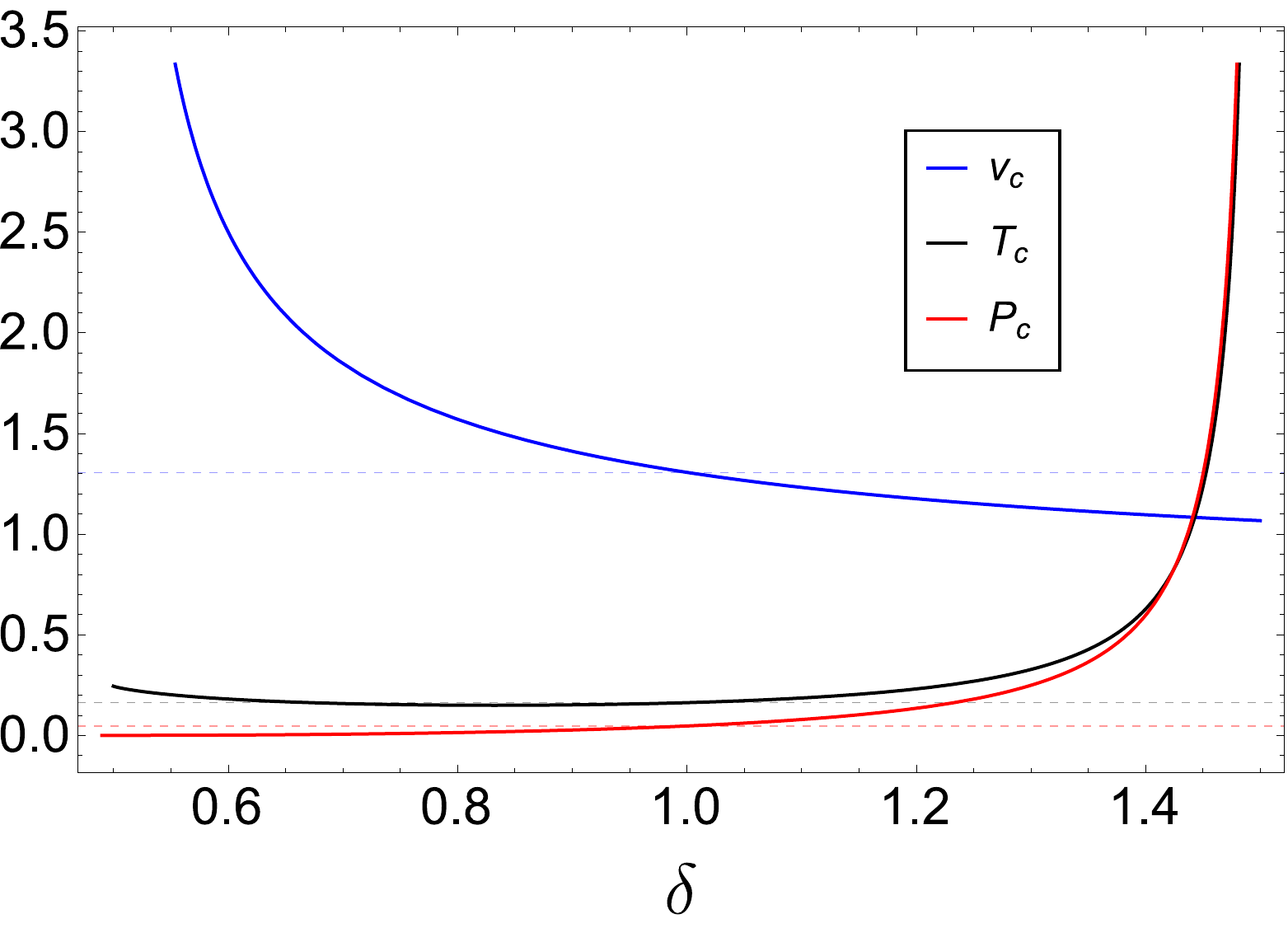}
\caption{Plot of $v_c$, $T_c$ and $P_c$ as a function of $\delta$ (online colors).}
\label{Fig4}
\end{center}
\end{figure}

The critical point of phase transitions can be found through the conditions~\eqref{C1}-\eqref{C2}, which now give
\begin{eqnarray}
\label{vc}
\hspace{-2mm}v_{c}&=&\frac{2 Q \left(1+2\delta\right)^{\frac{1}{2}}}{\left(\delta-\frac{1}{2}\right)^{\frac{1}{2}}\left(1-\eta^2\right)}\,,\\[2mm]
\label{Tc}
\hspace{-2mm}T_{c}&=&\frac{2^{\frac{1-2\delta}{2}}\pi^{-\delta}\left[Q\left(\frac{2\delta+1}{2\delta-1}\right)^{\frac{1}{2}}\right]^{1-2\delta}\left(1-\eta^2\right)^{\delta}}
{\delta\left(3-2\delta\right)\left(1+2\delta\right)}\,,\\[2mm]
\label{Pc}
\hspace{-2mm}P_{c}&=&\frac{\left(1-2\delta\right)^2\left(1-\eta^2\right)^2}
{32\pi Q^2\left(3-2\delta\right)\left(1+2\delta\right)}\,\,\Longrightarrow\,\, l_c=\left(\frac{3}{8\pi P_c}\right)^{\frac{1}{2}}. 
\end{eqnarray}
The critical radius $r_c=v_c/2$ corresponds  to the critical thermodynamic volume
\begin{eqnarray}
\nonumber
V_c&=&\left(\frac{\partial M}{\partial P}\right)_S\bigg|_{r_h=r_c}
=\frac{4}{3}\pi \left(1-\eta^2\right) r_c^3\\[2mm]
&=&\frac{4\pi Q^3}{3\left(\frac{2\delta-1}{4\delta+2}\right)^{\frac{3}{2}}\left(1-\eta^2\right)^2}\,.
\label{Volcrit}
\end{eqnarray}

Some comments are in order here: first,  we notice that the critical specific volume $v_c$ and temperature $T_c$ are mathematically well-defined, provided that $\delta>1/2$, while physically allowed (i.e. positive) temperature and pressure require $\delta<3/2$.\footnote{It is interesting to note that, for three-dimensional BHs, the value $\delta=3/2$ implies $S\propto A_{bh}^{3/2}\propto r_h^3$, corresponding to extensivity of Tsallis entropy~\cite{TsallisCirto}. More generally, for $d$-dimensional BHs, this is true when $\delta=d/(d-1)$.} The  standard expressions for AdS BHs with global monopole are correctly recovered in the $\delta\rightarrow1$ limit~\cite{Deng:2018wrd}. Furthermore, from Fig.~\ref{Fig4}
we notice that $v_c$ increases (decreases) for $\delta<1$ ($\delta>1$) respect to the corresponding $\delta=1$ value (which is marked by the horizontal dashed line), while the opposite behavior is exhibited
by $P_c$. On the other hand, $T_c$ lies above
the standard value for large and small Tsallis parameter,
while it drops below in the intermediate regime. Finally, we see that  $v_c\rightarrow0$, while $T_c,P_c\rightarrow\infty$ for $Q\rightarrow0$ regardless
of $\delta$, which entails that phase transitions are still peculiar to charged BHs in Tsallis theory, at least in pure Einstein gravity (by contrast, in~\cite{GB} it has been
argued that in Gauss-Bonnet gravity, BHs undergo small-large  transitions akin to those described above in the uncharged case too).

It is worth noting that the critical coefficient $P_c v_c/T_c$ now
obeys
\be
\label{coef}
\frac{P_c v_c}{T_c}=\frac{3}{8}\,\frac{\left(2\pi Q^2\right)^{\delta-1}\delta\left(2\delta-1\right)^{2-\delta}\left(1+2\delta\right)^\delta\left(1-\eta^2\right)^{1-\delta}}
{3}\,.
\ee
For $\delta=1$, the second fraction on the r.h.s. tends to unity and
the standard expression $P_c v_c/T_c=3/8$
for Reissner-Nordstr\"om AdS BHs is reproduced (see the discussion below Eq.~\eqref{cVan}).
Furthermore, performing the transformation
\be
p=\frac{P}{P_c}\,,\quad \nu=\frac{v}{v_c}\,,\quad \tau=\frac{T}{T_c}\,,
\ee
and making use of Eq.~\eqref{coef}, the equation of state~\eqref{eqphys}
can be cast as
\be
\label{lcs}
8\tau=3\nu\left(Ap+\frac{2 B}{\nu^2}\right)-\frac{D}{\nu^3}\,,
\ee
where
\begin{eqnarray}
A&\equiv&\frac{\nu^{2(1-\delta)}\left(4\delta^2-1\right)}{3}\,,\\[2mm]
B&\equiv&\frac{\nu^{2(1-\delta)}\left[3-4\delta\left(\delta-1\right)\right]}{3}\,,\\[2mm]
D&\equiv&-\nu^{2(1-\delta)}\left[3+4\delta\left(\delta-2\right)\right].
\end{eqnarray}
The relation~\eqref{lcs} is similar to the law of corresponding states for fluids~\cite{Kubiznak}
\begin{equation*}
8\tau=3\nu\left(p+\frac{2}{\nu^2}\right)-\frac{1}{\nu^3}\,.
\end{equation*}
The latter is consistently recovered for $\delta=1$, since $A=B=D=1$ in this limit.

\begin{figure}[t]
\begin{center}
\includegraphics[width=8.3cm]{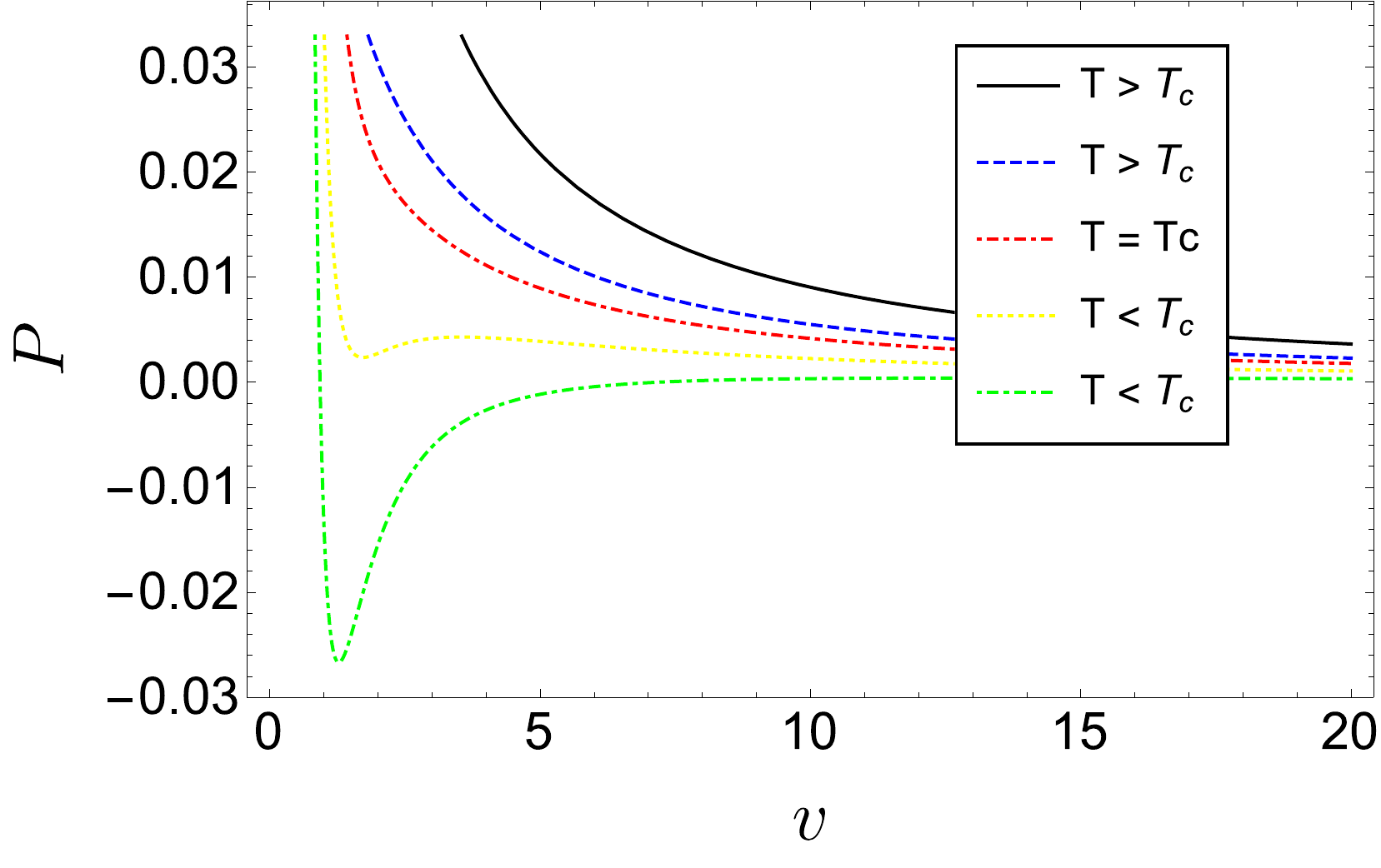}
\\[0.5cm]
\hspace{4mm}\includegraphics[width=8.0cm]{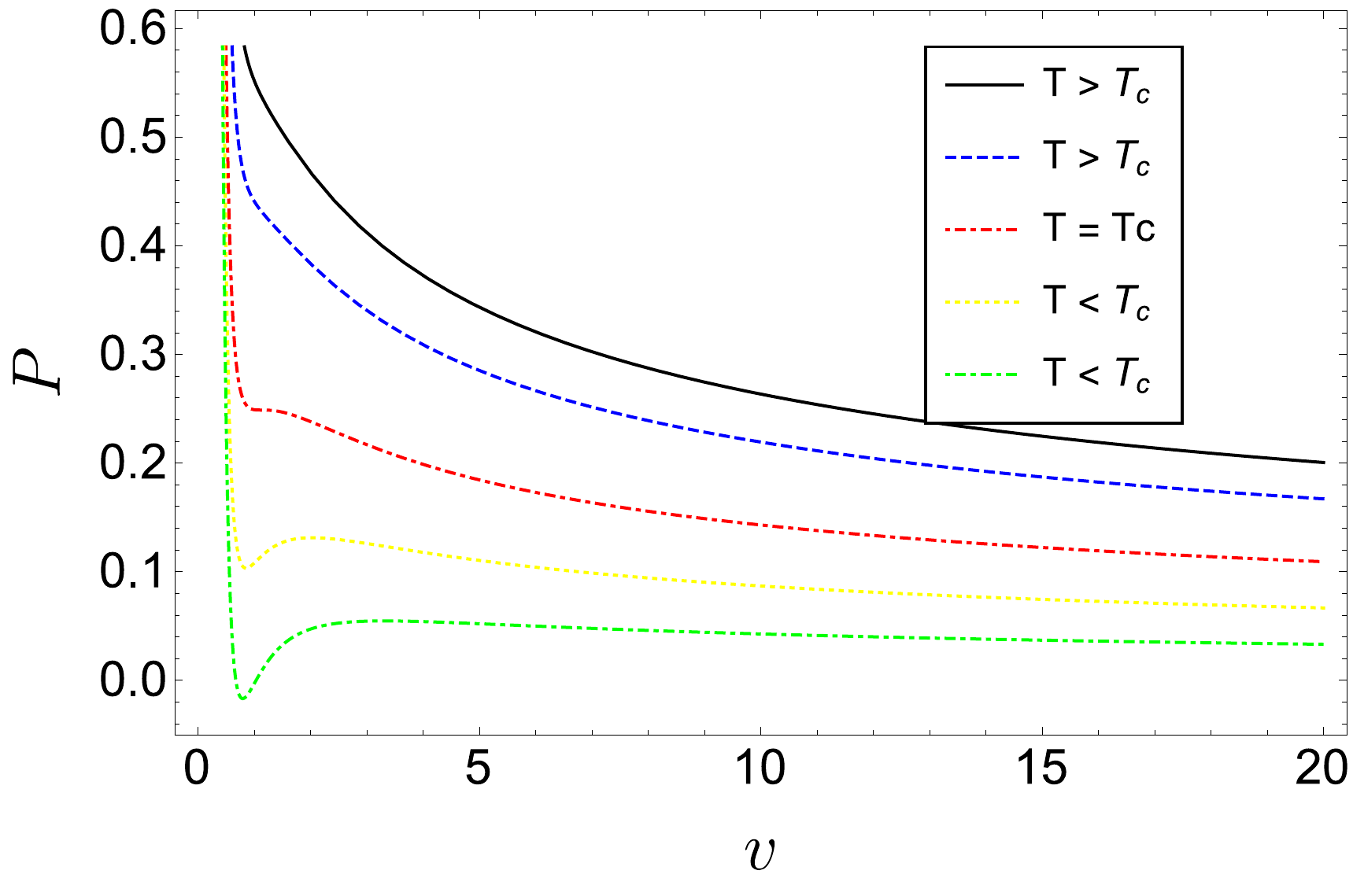}
\caption{$P-v$ diagram of charged AdS black holes with global monopole for $\delta=0.8$ (top panel) and $\delta=1.3$ (bottom panel). The temperature of the isotherms decreases from top to bottom. The red dot-dashed line indicates the critical isotherm at $T=T_c$ (online colors).}
\label{Fig5}
\end{center}
\end{figure}

To further elaborate on the connection between BHs and
van der Waals fluid, let us analyze the $P-v$ diagrams in the extended
phase space. These are displayed in Fig.~\ref{Fig5}, 
where it is shown that
AdS BHs with global monopole in Tsallis statistics qualitatively behave like a van der Waals liquid-gas system, with Tsallis parameter affecting the critical values $P_c, v_c$ and $T_c$. As observed below Eq.~\eqref{Cp}, a small-large BH phase transition occurs, which resembles the liquid-gas phase transition of van der Waals fluids (see Fig.~\ref{Fig0} for comparison). 
As far as $T<T_c$, each isotherm exhibits van der Waals-like oscillation, having a local minimum and maximum.
For $T=T_c$ the oscillating part squeezes and the two stationary points converge into the critical (inflection) point $(P_c,v_c, T_c)$ (see Eqs.~\eqref{vc}-\eqref{Pc}).
By further increasing the temperature to $T>T_c$, 
the inflection point disappears and $P$ decreases monotonically along each isotherm. Finally,
as explained below Eq.~\eqref{pv}, there is a threshold temperature $T_0$ below which the pressure becomes negative for certain values of $v$. Clearly, this branch is unphysical
and once the Maxwell's equal area is used, it
is replaced by the corresponding isobar of positive pressure.

\subsection{Gibbs free energy and global stability }
\label{GibbsSection}

Above we have studied Tsallis entropy effects on
the local thermodynamical properties of charged AdS BHs. 
We now proceed to analyze
the global stability by investigating Gibbs free energy. Toward this end,
we use Eq.~\eqref{Gibbsdif}, which after integration can be written for BHs
as~\cite{Deng:2018wrd}
\begin{eqnarray}
\label{Gibbs}
G(T,P)&=&M-TS\\[2mm]
&&\hspace{-19mm}=\,\frac{Q^2\left(3+6\delta\right)+r_h^2\left[-3+6\delta+8 \pi\hspace{0.2mm} P \hspace{0.2mm} r_h^2\left(2\delta-3\right)\right]\left(1-\eta^2\right)^2}
{12\delta r_h\left(1-\eta^2\right)}.
\nonumber
\end{eqnarray}
Here, $r_h$ should be understood as a function of $P$ and $T$, i.e. $r_h\equiv r_h(P,T)$, via the equation of state~\eqref{eqphys}.

\begin{figure}[t]
\begin{center}
\includegraphics[width=8.3cm]{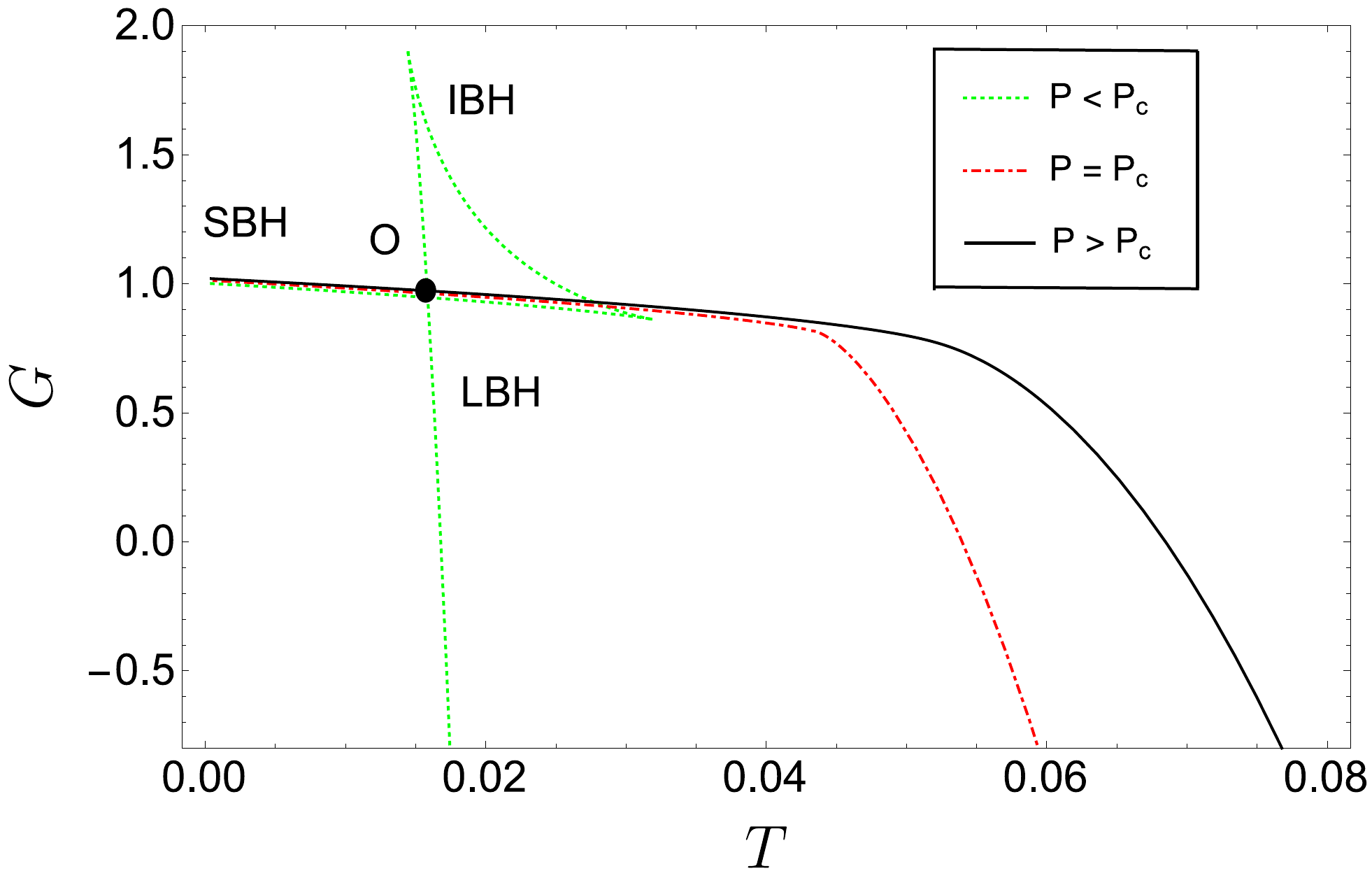}
\caption{Plot of $G$ as a function of $T$ for $\delta=0.8$ (the same qualitative behavior is obtained for other $\delta$'s consistent with Eqs.~\eqref{vc}-\eqref{Pc}). The red dot-dashed line indicates the critical isobar at $P=P_c$ (online colors).}
\label{Fig6new}
\end{center}
\end{figure}

The behavior of Gibbs free energy versus $T$ for three different values of $P$ is depicted in 
Fig.~\ref{Fig6new},  which is to be
compared with Fig.~\ref{Fig0bis}.
For $P<P_c$, $G$ exhibits the characteristic swallow tail behavior, which signals a first order phase transition in the system. In particular, BH first runs along the SBH
branch until $T$ increases to the critical point $O$. Then, both the
SBH and LBH coexist, since they share the
same Gibbs free energy. 

\begin{figure}[t]
\hspace{-14mm}\includegraphics[width=10.7cm]{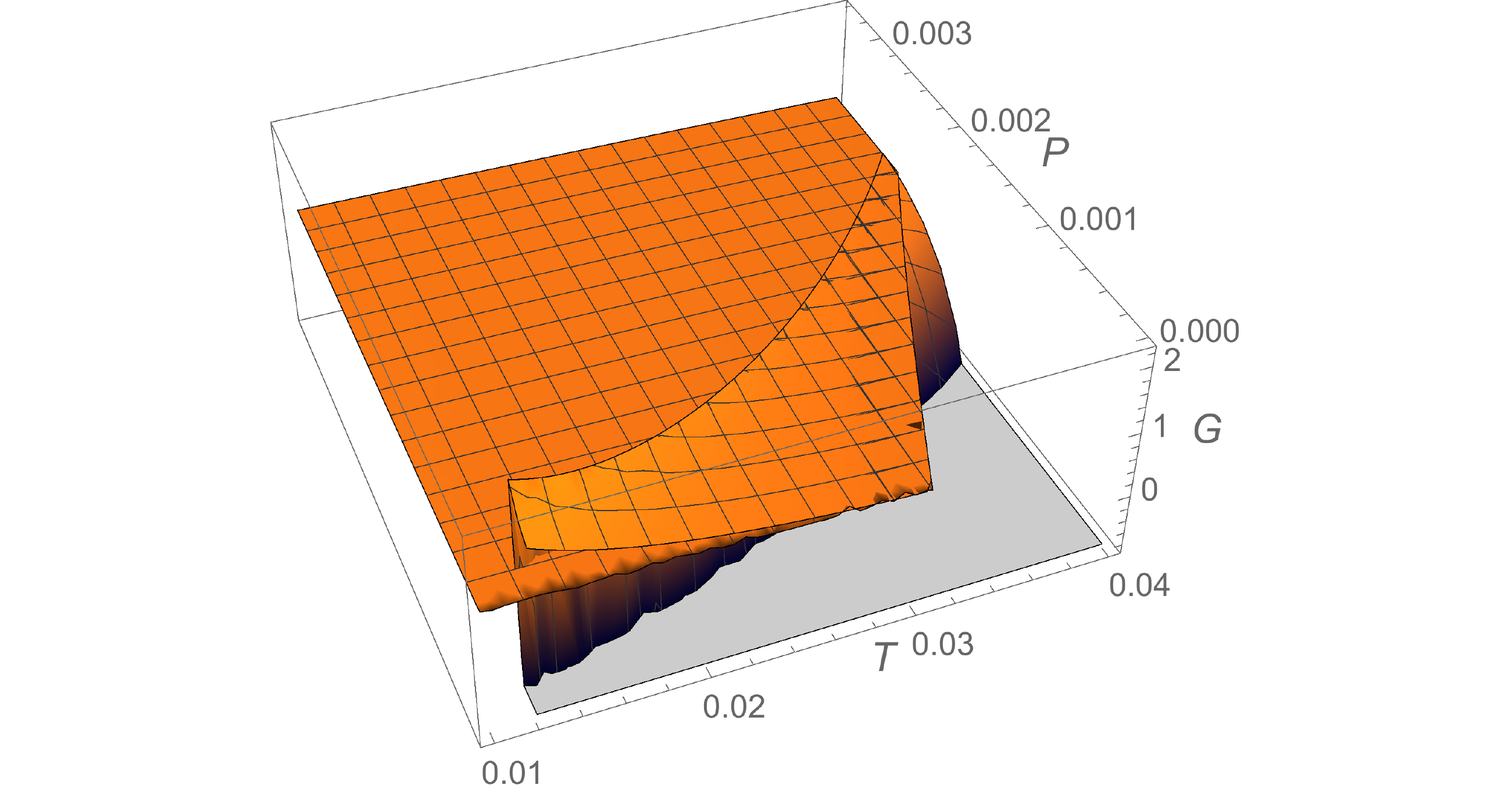}
\caption{3D plot of $G$ as a function of $P$ and $T$ for $\delta=0.8$ (the same qualitative behavior is obtained for other $\delta$'s consistent with Eqs.~\eqref{vc}-\eqref{Pc}). The figure shows the characteristic swallow tail behavior of Gibbs free energy, which corresponds to the first-order small-large BH transition occurring at the intersection of $G$ surfaces. Also, the coexistence line is visible.}
\label{Fig3D}
\end{figure}

The coexistence line in the $P-T$ plane
can be obtained by finding a curve for which the Gibbs free energy and temperature are equal for SBH and LBH. In so doing, one obtains the same curve as in Fig.~\ref{Fig0ter}, where the role of the liquid and gas phases is now played by 
the small and large ``phases'' of BHs, respectively. 
The coexistence line is evident from the cross-section of the 3D plot in Fig.~\ref{Fig3D}.  As the temperature further increases, LBH becomes statistically favored respect to the small and intermediate black hole (IBH) stages because of its lower Gibbs free energy. The point $O$ marks the small-large BH first-order phase transition. We also notice that, because of Eq.~\eqref{TsaEN}, such transition corresponds to a discontinuity of the entropy. In turn, this means that there is a release of latent heat\footnote{The small-to-intermediate and intermediate-to-large BH transitions, where heat capacity and second order derivatives of $G$ are discontinuous  but entropy and volume remain continuous, should correspond to second-order phase transition points according to Ehrenfest classification. Such transitions for BHs have been first discussed by Davies in~\cite{Davies1,Davies2}. Contrary to first-order phase transitions, however, the physics underlying BH second-order transitions seems to be still rather obscure. For more details on the subject, see~\cite{Sokolowski:1980uva,Banerjee}.}.

\subsection{Critical parameters}
\label{critpar}

Following the discussion at the end of Sec.~\ref{vdwSec},
let us now derive the basic critical exponents $\alpha,\beta,\gamma$ and $\Delta$ for AdS BHs. 
We start by considering the specific heat at
constant thermodynamic volume $C_V=T\left(\frac{\partial S}{\partial T}\right)_V$. To compute this variable,
we introduce the free energy
\begin{eqnarray}
&&\hspace{-3mm}F(T,V)=G-PV\\[2mm]
&&\hspace{-3mm}=\,\frac{Q^2-r_h\left(1-\eta^2\right)\left\{2r_h^{2\delta}\hspace{0.2mm}T\left[\pi\left(1-\eta^2\right)\right]^\delta-r_h\left(1-\eta^2\right)\right\}}
{2r_h\left(1-\eta^2\right)}\,.
\nonumber
\end{eqnarray}
Hence, the entropy is
\begin{equation*}
S(T,V)=-\left(\frac{\partial F}{\partial T}\right)_V=\left[\pi\left(1-\eta^2\right)r_h^2\right]^{\delta}\,,
\end{equation*}
which gives Eq.~\eqref{TsEn}, as expected. In turn, we find $C_V=0$, which
implies $\alpha=0$ from Eq.~\eqref{alpha}.

Concerning $\beta$, let us approximate the
generalized law of corresponding states~\eqref{lcs}
around a critical point as in Eq.~\eqref{TildeQ}. 
Toward this end, we define
\be
\tilde T=\frac{T}{T_c}-1\,, \qquad \omega=\frac{V}{V_c}-1\,,
\ee
which lead to
\begin{eqnarray}
\nonumber
p&=&1+\frac{8}{4\delta^2-1}\hspace{0.2mm}\tilde T+\frac{8\left(2\delta-3\right)}{4\delta^2-1}\hspace{0.2mm}\tilde T\omega+\left(\frac{8\delta}{3}-4\right)\omega^3\\[2mm]
&&+\,\mathcal{O}(\tilde T\omega^2,\omega^4)\,.
\label{papp}
\end{eqnarray}
The truncation of this expansion is justified
by Eq.~\eqref{Justif} below.  As noticed in~\cite{Kubiznak},
one could equivalently work in terms of
$\omega'=\frac{v}{v_c}-1$ and obtain
the same results for the critical exponents.

If we now differentiate Eq.~\eqref{papp} for a fixed $\tilde T<0$,
we obtain
\begin{eqnarray}
\nonumber
dP&=&P_c\, dp\\[2mm] 
&=& P_c\left[\frac{8\tilde T\left(2\delta-3\right)}{4\delta^2-1}+3\left(\frac{8\delta}{3}-4\right)\omega^2\right]d\omega\,.
\end{eqnarray}
Using Maxwell's equal area law\footnote{It is known that Maxwell's equal area law allows to describe phase transitions of fluids by replacing the oscillating part of the isotherms with $T<T_c$ by an isobar, according to Eq.~\eqref{Max}. This prescription reflects the property that the two phases at the transition share the same Gibbs free energy. The same considerations can be extended to BH transitions by identifying the liquid and gas phases with small and large stages of BHs, respectively.}
\be
\label{Max}
\oint v dP =0\,,
\ee
and the fact that the pressure
remains constant during the phase transition, we arrive to the following two equations
\begin{eqnarray}
\nonumber
&&\left[1+\frac{8}{4\delta^2-1}\hspace{0.2mm}\tilde T+\frac{8\left(2\delta-3\right)}{4\delta^2-1}\hspace{0.2mm}\tilde T\omega+\left(\frac{8\delta}{3}-4\right)\omega^3\right]_{\omega=\omega_l}\\[2mm]
&=&\left[1+\frac{8}{4\delta^2-1}\hspace{0.2mm}\tilde T+\frac{8\left(2\delta-3\right)}{4\delta^2-1}\hspace{0.2mm}\tilde T\omega+\left(\frac{8\delta}{3}-4\right)\omega^3\right]_{\omega=\omega_s}
\label{1eq}
\end{eqnarray}
and
\be
\label{2eq}
0=\int_{\omega_l}^{\omega_s}\omega \left[\frac{8\tilde T\left(2\delta-3\right)}{4\delta^2-1}+3\left(\frac{8\delta}{3}-4\right)\omega^2\right]d\omega
\ee
where we have denoted by $\omega_{s,l}$
the `volume' of small and large BHs, respectively.

It is easy to verify that the unique non-trivial root of 
Eqs.~\eqref{1eq}-\eqref{2eq} that recovers the correct behavior
in the $\delta=1$ limit~\cite{Kubiznak} is 
\be
\omega_s=-\omega_l=\left[-\frac{6\tilde T}{\left(2\delta-1\right)\left(1+2\delta\right)}\right]^{\frac{1}{2}}\,,
\ee
where we are supposing $\delta>1/2$, consistently
with the discussion below Eq.~\eqref{Volcrit}.
Thus, from Eq.~\eqref{beta} we find
\be
\label{Justif}
\eta=V_c\left(\omega_l-\omega_s\right)=2V_c\omega_l\propto \sqrt{-\tilde T}\,,
\ee
which implies $\beta=1/2$.

Next, to calculate $\gamma$ we differentiate Eq.~\eqref{papp}
to obtain
\be
\left(\frac{dV}{dP}\right)_T=-\frac{4\delta^2-1}{8(3-2\delta)}\hspace{0.2mm}
\frac{V_c}{P_c}\hspace{0.2mm}\frac{1}{\tilde T}+\mathcal{O}(\omega^2)\,.
\ee
Therefore, it follows that
\be
\kappa_T=-\frac{1}{V}\hspace{0.2mm}\left(\frac{\partial V}{\partial P}\right)_T\propto \frac{1}{\tilde T}\,,
\ee
which entails $\gamma=1$ from Eq.~\eqref{gamma}.

We also notice that, at the critical point, only $C_p$
diverges, while $C_V$ remains finite. Indeed,
from Eqs.~\eqref{alpha},~\eqref{TempS} and~\eqref{defCP} 
we have
$C_V=0$ and
\begin{eqnarray}
C_p&=&\\[2mm]
\nonumber
&&\hspace{-10mm}\frac{2S\delta\left[\pi Q^2-S^{\frac{1}{\delta}}\left(8P S^{\frac{1}{\delta}}+1-\eta^2\right)\right]}{S^{\frac{1}{\delta}}\left[8PS^{\frac{1}{\delta}}\left(2\delta-3\right)+\left(2\delta-1\right)\left(1-\eta^2\right)\right]-\pi Q^2\left(2\delta+1\right)}\,.
\end{eqnarray}
This is singular for
\begin{eqnarray}
\label{Ssing}
S=
\left\{\frac{\left(1-2\delta\right)\left(1-\eta^2\right)\pm f(P)^{\frac{1}{2}}}
{16P\left(2\delta-3\right)}\right\}^{\delta}\,,
\end{eqnarray}
where
\be
f(P)\equiv32\pi Q^2 P\left(2\delta-3\right)\left(1+2\delta\right)+\left(1-2\delta\right)^2\left(1-\eta^2\right)^2\,.
\ee
By using Tsallis entropy~\eqref{TsEn},
from Eqs.~\eqref{defvrh} we get two solutions for $v$ corresponding to the entropy~\eqref{Ssing}. Of these, the physical one is
\be
v=\frac{1}{2}\left[\frac{\left(1-2\delta\right)\left(1-\eta^2\right)+f(P)}
{\pi P\left(1-\eta^2\right)\left(2\delta-3\right)}\right]^{\frac{1}{2}}\,.
\ee
It is straightforward to check that this equation is exactly
satisfied at the critical point~\eqref{vc}-\eqref{Pc}, where $f(P)=0$.
In this way, we recover the  result of Sec.~\ref{CritP}.

Finally, the shape of the critical isotherm $\tilde T=0$ is fixed by Eq.~\eqref{papp}
\be
p-1=\left(\frac{8\delta}{3}-4\right)\omega^3\,,
\ee
which gives $\Delta=3$.
This finalizes the computation of basic critical exponents.

From comparison with results
in Sec.~\ref{vdwSec}, we infer that BHs with non-additive
Tsallis entropy exhibit a van der Waals fluid-like
behavior not only at qualitative, but also quantitative level.

\subsection{Sparsity}
Sparsity is a relevant feature characterizing BH radiation. 
It is defined as the average time between the
emission of successive quanta over the timescales set by
the energies of the emitted quanta. One of the most important
aspects of Hawking radiation is that it
appears extremely sparse during the evaporation process 
as compared to black body radiation~\cite{Spars}, thus allowing
to differentiate between the two phenomena. 

Sparsity can be quantified 
by introducing the parameter\footnote{Here we denote the sparsity parameter by $\tilde\eta$ instead of the usual $\eta$ in order not to create confusion with the monopole parameter.}
\be
\label{teta}
\tilde \eta=\frac{C}{g}\left(\frac{\lambda_t^2}{A_{eff}}\right),
\ee
where $C$ is a dimensionless constant, $g$ the spin
degeneracy factor of the emitted quanta, $\lambda_t=2\pi /T$
the thermal wavelength and $A_{eff}=27A_{bh}/4$ the effective
area of the BH. For the case of Schwarzschild BHs and massless bosons, one has $\lambda_t=2\pi/T_H=8\pi^2r_h$, which in turns implies 
\be
\tilde \eta_H=\frac{64\pi^3}{27}\approx 73.49\,, 
\ee
which is constant and much higher than unity (for comparison, we remark that $\tilde \eta\ll1$ for black body radiation). 

In~\cite{Alonsobis} the impact of generalized models of Heisenberg relation has been studied on Eq.~\eqref{teta}, 
showing that sparsity decreases toward the late stages of evaporation process. Similarly, in~\cite{Alonso} the sparsity of 
R\'enyi radiation has been considered in the context of non-extensive statistics. The interesting result is that such a radiation
is initially not sparse, however as evaporation progresses, 
$\tilde\eta$ increases and eventually approaches the standard value $\tilde\eta_H$.

The above studies concern the behavior of sparsity for  Schwarzschild BHs. Here, we intend to extend these considerations
to the case of charged AdS BHs with global monopole within Tsallis entropy-based framework. Toward this end, we apply the definition~\eqref{teta} with $T$ being given by Eq.~\eqref{TempS}. After some algebra, we get
\be
\tilde \eta_\delta (S)= \tilde\eta_H\frac{\pi^{2}l^4\delta^2\left(1-\eta^2\right)^2S^{2}}
{\left\{\pi l^2\left[\pi Q^2-S^{\frac{1}{\delta}}\left(1-\eta^2\right)\right]-3S^{\frac{2}{\delta}}\right\}^2}\,.
\ee 
First, we observe that for $Q=\eta=0$ and sufficiently large AdS radius $l$, the above expression
correctly reduces to the sparsity of Tsallis radiation obtained
in~\cite{Cimidiker:2023kle} for the case of Schwarzschild BHs. 
Also, the additional $\delta\rightarrow1$ limit gives $\tilde\eta_\delta=\tilde\eta_H$, as expected. Apparently, $\tilde\eta_\delta$ diverges
for 
\begin{equation*}
\label{divsp}
S=\left\{\frac{\pi}{6}\left[\left[12l^2Q^2+l^4\left(1-\eta^2\right)^2\right]^{\frac{1}{2}}-l^2\left(1-\eta^2\right)\right]\right\}^\delta\,.
\end{equation*}
However, this value corresponds to the physical limitation point $S_0$ defined in Eq.~\eqref{plp}. Therefore, in order for $T$ to be positively defined, we have to restrict our attention 
to the interval $S>S_0$.

\begin{figure}[t]
\begin{center}
\includegraphics[width=8.3cm]{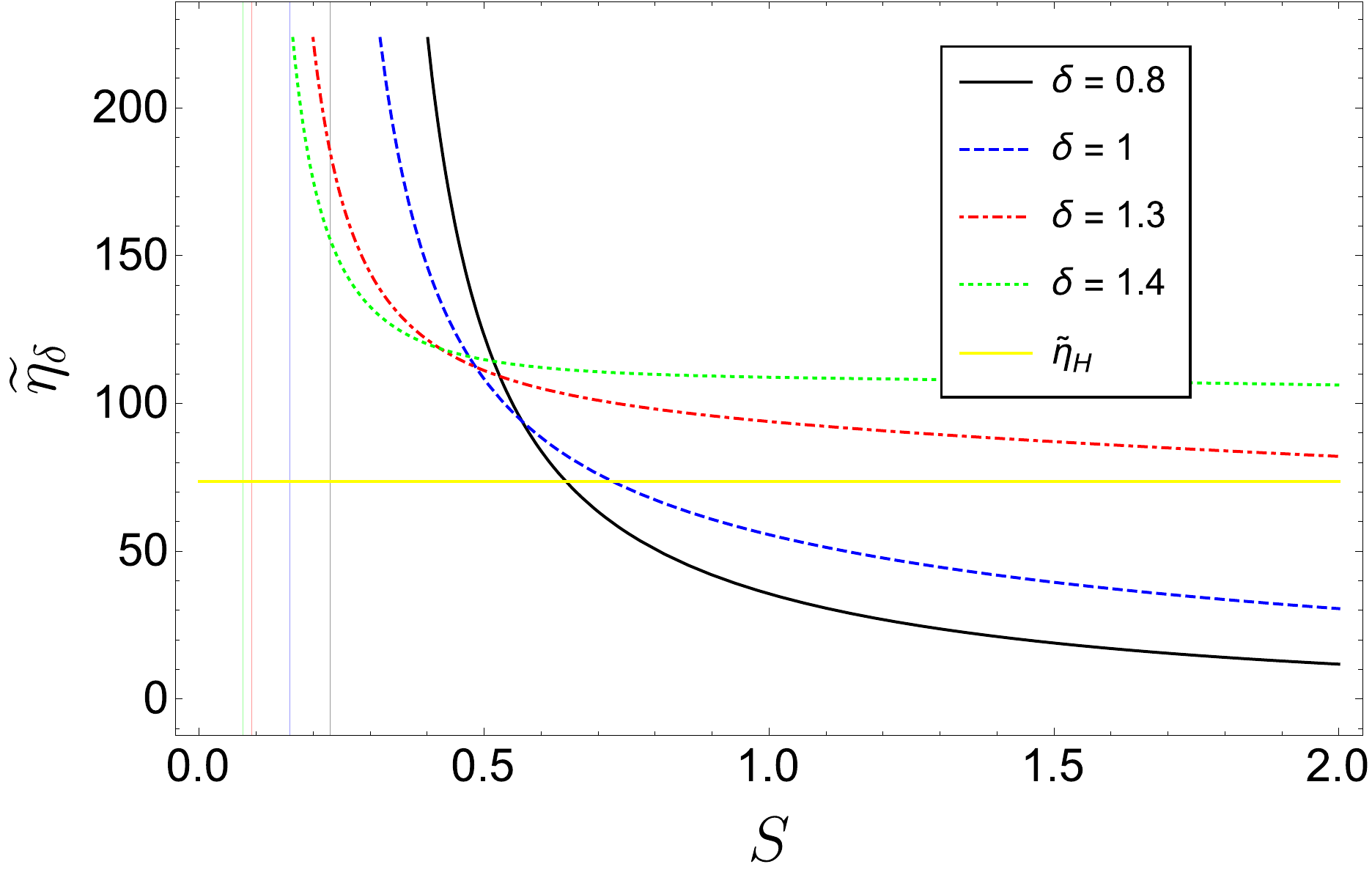}
\caption{Plot of $\tilde\eta_\delta$ as a function of $S$ for various values of $\delta$ and $l=2$. The vertical lines represent the physical limitation entropy $S_0$ for each curve (online colors).}
\label{spars}
\end{center}
\end{figure}

The behavior of $\tilde\eta_\delta$ as a function of $S$ is plotted in Fig.~\ref{spars} for various values of $\delta$ and fixed $Q,\eta,l$. The yellow solid line representing $\tilde\eta_H$ is displayed for comparison. We can see that BH sparsity is significantly affected in the present analysis. In particular, 
it lies above the standard value $\tilde\eta_H$ for small $S$, 
which means that radiation is more sparse than Hawking case at this stage of BH evolution.
As the entropy increases, $\tilde\eta_\delta$ 
decreases monotonically and vanishes asymptotically. 
This indicates that BH radiation is not sparse at all in this regime and
nearly behaves like a black body radiation. 
Tsallis exponent affects the rate of variation of sparsity.
Indeed, for $\delta<1$ (black solid curve) $\tilde\eta_\delta$ asymptotically decreases to zero faster than the ordinary $\delta=1$ sparsity (blue dashed curve), while the opposite behavior occurs  for $\delta>1$. 

\section{Geometrothermodynamics of charged black holes}
\label{Geom}

Let us now study BH thermodynamics from a geometric perspective.  Toward this end, we remark that first Weinhold and later Ruppeiner introduced a geometric framework to explore the microscopic behavior of thermodynamic systems. Specifically, they defined a Riemannian metric as the second derivative of internal energy~\cite{Wein1} and entropy~\cite{Rupp1}  with respect to given thermodynamic variables, respectively. Thus, both these two approaches provide an effort to extract the microscopic interaction information from the axioms of thermodynamics phenomenologically or qualitatively.

In the above frameworks, the sign of the  scalar curvature $R$ of the metric is directly related to the nature of  microscopic interactions, with  negative curvature corresponding to prevailingly attractive forces and positive curvature to repulsive ones. Furthermore, the curvature vanishes for systems where interactions are perfectly balanced. We remark that this picture inherits the considerations first emphasized in~\cite{Oshima}
for the case of Fermi ($R>0$) and Bose ($R<0$) ideal gas and classical ideal ($R=0$) gas. Though it is known that Weinhold and 
Ruppeiner formalisms still have the status of conjecture, the 
correspondence between the sign of the curvature and the behavior of microinteractions has been verified for a large number of statistical model so far~\cite{Rupp2,Rupp3}.

Since one can define a thermodynamic for BHs, it is quite natural to think of an associated microscopic structure of interacting constituents. Along this line, the theory of thermodynamic
geometry has also been applied to BHs as an effective perspective for studying the micro-mechanism of sub-interactions phenomenologically. This has been done,  for example, in~\cite{Cai:1998ep,Wei:2015iwa,Wei:2019uqg,Xu:2020gud,Ghosh:2020kba}.

In the mass representation, Weinhold metric is specified as~\cite{Wein1,Soroushfar:2020wch} 
\be
g_{ij}^w=-\partial_i\partial_j M(S,P,Q)\,.
\ee
In turn, the related line element is
\be
ds^2_w=g^w_{ij} dx^i dx^j\,,
\ee
where we have we have generically denoted the 
independent fluctuation coordinates by $x^i$.

On the other hand,  in Ruppeiner formalism one considers 
the system entropy as its thermodynamic potential, yielding 
\be
\label{R1}
g_{ij}^{Rup}=-\partial_i\partial_j S\,.
\ee
Based on this equation, 
it is straightforward to show that the line elements of Weinhold and 
Ruppeiner metrics differ only by a conformal factor~\cite{Mrugala}, that is
\be
\label{R2}
ds^2_R=\frac{ds^2_w}{T}\,.
\ee 

\subsection{AdS black holes}

Following~\cite{Wei:2015iwa}, here we focus on 
Ruppeiner formalism, which 
is expected to be more powerful to decipher
the character of interactions among BH micromolecules. 
By treating, for example, the entropy and pressure as the fluctuation coordinates, the Ruppeiner  scalar curvature  $R^{Rup}$ for charged AdS BHs with a global monopole on the $S-P$ plane reads 
\be
R^{Rup}(S,P)=\frac{2\pi Q^2-S^{\frac{1}{\delta}}\left(1-\eta^2\right)}{\delta \,S\left[S^{\frac{1}{\delta}}\left(8P S^{\frac{1}{\delta}}+1-\eta^2\right)-\pi Q^2\right]}\,.
\ee
Notice that this expression reduces to the curvature computed in~\cite{Guo:2019oad} for $\eta\rightarrow0$ and $\delta\rightarrow1$, as expected.  

\begin{figure}[t]
\begin{center}
\includegraphics[width=8.2cm]{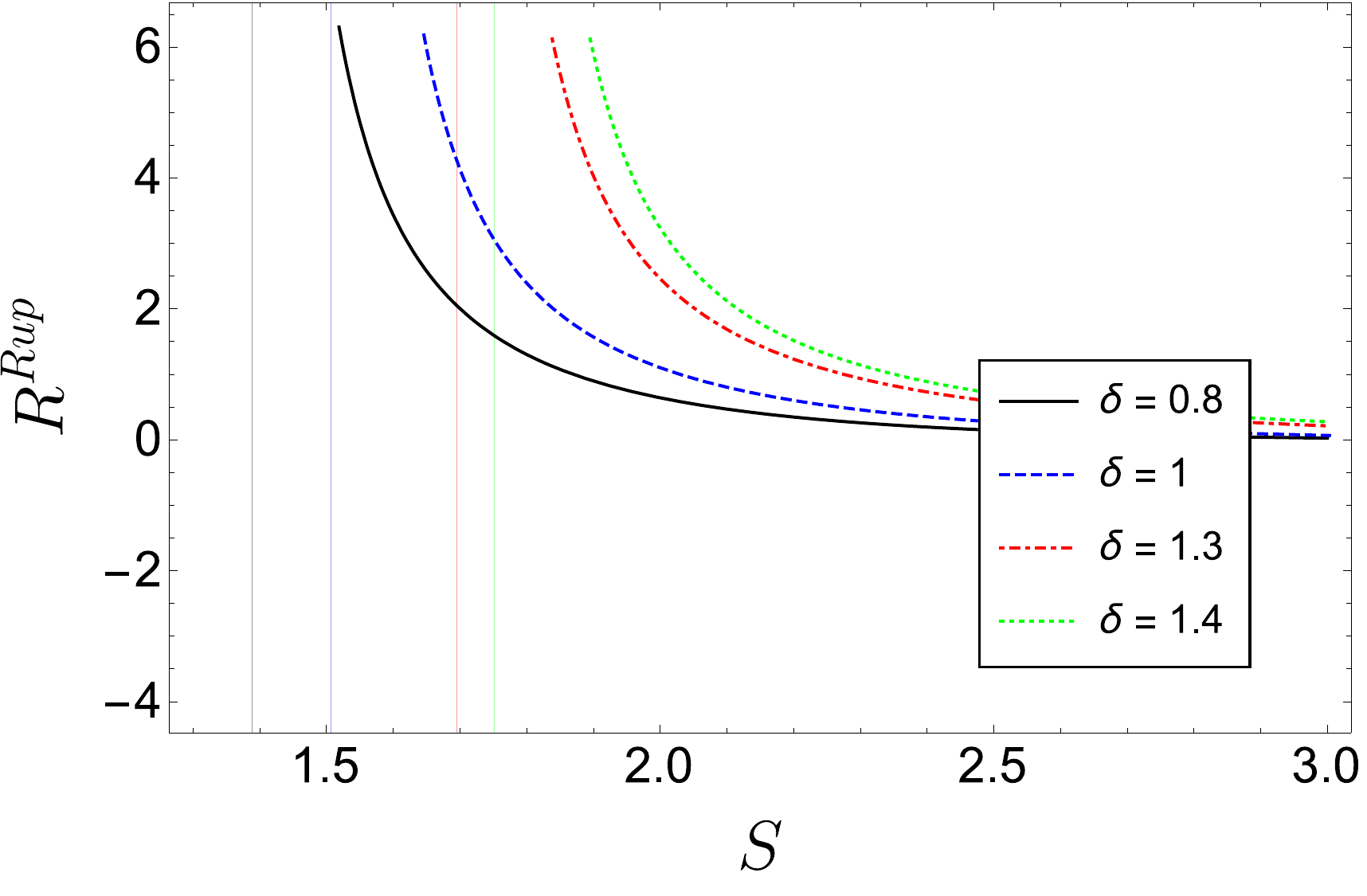}
\caption{Plot of $R^{Rup}$ as a function of $S$ for various values of $\delta$. The parameter $l$ is fixed to $l= 10 l_{c}$ defined through the critical condition~\eqref{Pc}. The vertical lines represent the physical limitation entropy $S_0$ for each curve
(online colors).}
\label{FigRup}
\end{center}
\end{figure}

The behavior of $R^{Rup}$ versus
the entropy $S$ is displayed in Fig.~\ref{FigRup} 
for various values of $\delta$ and fixed $Q=0.6, \eta=0.5$. 
We see that $R^{Rup}$ exhibits one singularity, 
which coincides with the physical limitation point $S_0$.
Thus, as discussed for sparsity, we only consider values of entropy higher than this threshold. For sufficiently small $S$, we have $R^{Rup}>0$, which indicates repulsive forces among charged microstructures of AdS BHs. This confirms the results of~\cite{Wei:2015iwa,Guo:2019oad,Wei:2019uqg,Wei:2019yvs,Ghosh:2020kba},
where the general statement has been claimed that the scalar curvature of BH systems with charged molecules is positive. 
Such a behavior vaguely resembles that of
van der Waals fluids, where the particles have a hard repulsive core and an infinitesimally small attraction of infinite range.
The crossing of Ruppeiner  curvature from positive to vanishing values as the entropy increases indicates that the initial repulsion progressively vanishes.
Asymptotically (i.e. for large BH horizon radii),
BH constituents end up being so far apart that $R^{Rup}\rightarrow0$, 
suggesting that BHs behave as effectively non-interacting objects in this regime. 
 
While not affecting the qualitative behavior of $R^{Rup}$,  it is observed that Tsallis prescription influences the entropy (i.e. radius) intervals where net intermolecular forces become manifest/disappear. For $\delta<1$ (black solid curve), the transition from positive  to vanishing curvature occurs at lower values of $S$ comparing to the standard scenario ($\delta=1$, blue dashed curve), which means that the initial configuration with dominant repulsion is disfavored respect to the non-interacting one, as compared to the $\delta=1$ case. This property resembles the behavior of systems with superadditive entropy, where the individual subconstituents of a given composite system (the BH microstructures in our case) tend to merge 
more strongly than the additive scenario~\cite{Vedral}. 
The above tendency is reversed for $\delta>1$,
thus revealing a subadditive-like nature of the entropy.

The physics underlying the aforementioned shifting of interactions
can be understood qualitatively in terms of the two fluid model~\cite{Guo:2019oad,Ghosh:2019pwy}. This system is presented as a binary mixture of two fluids with purely repulsive and attractive interactions among subconstituents, respectively.  The crossing point
is then fixed by the relative number densities of the molecules of the two fluids, which compete each other to determine the
prevailing behavior.  We also emphasize that, in the mean field
theory, the nature of microinteractions can be modeled by a suitable
potential. For the case of van der Waals fluid, this
is given by the Lennard-Jones potential, which turns out to be
repulsive at short distances and attractive in the long range.
Although attraction is dominant, repulsive forces are non-negligible
because of thermal fluctuations and/or molecular collisions.

As a final comment, we would like to mention that in~\cite{Soroushfar:2020wch} BH phase transitions have been investigated by considering the geometric structure of Quevedo (case-I and case-II) and Hendi-Panahiyan-Eslam-Momennia (HPEM) formalisms, besides Weinhold
and Ruppeiner approaches. 
Remarkably, it has been shown that, while the scalar curvature of Weinhold and Ruppeiner metrics has one singular point coinciding with the physical limitation point - in agreement with our result - the  scalar curvature of Quevedo metric (case-I) has three singularities, which correspond to the zero point (physical limitation) and divergence (transition critical) points of heat capacity, respectively. On the other hand, two of the three singularities  of the  scalar curvature of Quevedo (case-II) metric coincide
with the transition critical points of heat capacity, while no point corresponds to the zero point of heat capacity. Finally, in the case of HPEM metric,  the divergence points of the Ricci scalar coincide with the zero point and divergence points of heat capacity, respectively. 
As a perspective, it would be interesting to 
develop similar computations in the present framework
to see whether results of~\cite{Soroushfar:2020wch} are affected by non-additive entropy.

\subsection{BTZ black holes}
\label{BTZ}
Let us now apply the above geometrothermodynamic considerations
to $(2 +1)$-dimensional charged (non-rotating) 
BTZ BHs. While being easier to handle mathematically,  such BHs still provide a useful toy model to shed light into the behavior of BHs in $d\ge4$. 

The metric function for $(2 +1)$-dimensional charged  
BTZ BHs is given by~\cite{Martinez}
\be
f_{btz}(r)=-2m_{btz}+\frac{r^2}{l^2}-\frac{Q^2}{2}\log\left(\frac{r}{l}\right)\,,
\ee
with the horizon at $f(r_h)=0$. Furthermore, 
the relevant thermodynamic quantities read~\cite{Jawad:2022lww,Frassino:2015oca}
\begin{eqnarray}
\label{sbtz}
S_{btz}&=&\frac{\pi r_h}{2}\,,\\[2mm]
\nonumber
M_{btz}&=&\frac{m_{btz}}{4}=\frac{r_h^2}{8l^2}-\frac{Q^2}{16}\log\left(\frac{r_h}{l}\right)\\[2mm]
\label{Mbtz}
&=&\frac{S_{btz}^2}{2\pi^2l^2}-\frac{Q^2}{16}\log\left(\frac{2 S_{btz}}{\pi l}\right)\,,
\\[2mm]
T_{btz}&=&\frac{r_h}{2\pi l^2}-\frac{Q^2}{8\pi r_h}\,=\,\frac{S_{btz}}{\pi^2 l^2}-\frac{Q^2}{16S_{btz}}\,,\\[2mm]
P_{btz}&=&\frac{1}{8\pi l^2}\,=\,-\frac{\Lambda}{8\pi}\,,\\[2mm]
\label{vbtz}
V_{btz}&=&\pi r_h^2-\frac{\pi}{4}Q^2l^2\,=\,\frac{4}{\pi}S_{btz}^2-\frac{\pi}{4}Q^2l^2\,,\\[2mm]
\varphi_{btz}&=&-\frac{Q}{8}\hspace{0.2mm}\log\left(\frac{r_h}{l}\right)=-\frac{Q}{8}\log\left(\frac{2S_{btz}}{\pi l}\right)\,,
\label{phibtz}
\end{eqnarray}
where we have neglected global monopole effects for direct comparison with results of~\cite{Ghosh:2020kba}. Clearly, 
such effects can be accounted for by properly modifying
Eq.~\eqref{sbtz}-\eqref{phibtz} as done in Sec.~\ref{Pht}.

After implementation of Tsallis prescription~\eqref{TsaEN}, 
Eqs.~\eqref{sbtz}-\eqref{phibtz} become
\begin{eqnarray}
\label{91}
S_{btz}&=&\left(\frac{\pi r_h}{2}\right)^{\delta}\,,\\[2mm]
\label{mbtz}
M_{btz}&=&\frac{S_{btz}^{\frac{2}{\delta}}}{2\pi^2l^2}-\frac{Q^2}{16}\log\left(\frac{2 S_{btz}^{\frac{1}{\delta}}}{\pi l}\right),\\[2mm]
\label{Tbtz}
T_{btz}&=&\frac{S_{btz}^{\frac{2-\delta}{\delta}}}{\pi^2l^2\delta}-\frac{Q^2}{16\hspace{0.2mm}\delta\hspace{0.2mm} S_{btz}}\,,\\[2mm]
\label{94}
V_{btz}&=&\frac{4}{\pi}S_{btz}^{\frac{2}{\delta}}-\frac{\pi}{4}Q^2l^2\,,\\[2mm]
\varphi_{btz}&=&-\frac{Q}{8}\log\left(\frac{2S^{\frac{1}{\delta}}_{btz}}{\pi l}\right)\,.
\end{eqnarray}
We can now write the temperature and thermodynamic volume in the $S-P$ space as
\begin{eqnarray}
\label{planeT}
T_{btz}&=&\frac{8P_{btz}\hspace{0.5mm} S_{btz}^{\frac{2-\delta}{\delta}}}{\pi\delta}-\frac{Q^2}{16\delta S_{btz}}\,,\\[2mm]
V_{btz}&=&\frac{4}{\pi}S_{btz}^{\frac{2}{\delta}}-\frac{Q^2}{32P_{btz}}\,.
\label{planeV}
\end{eqnarray}

\begin{figure}[t]
\begin{center}
\includegraphics[width=8.2cm]{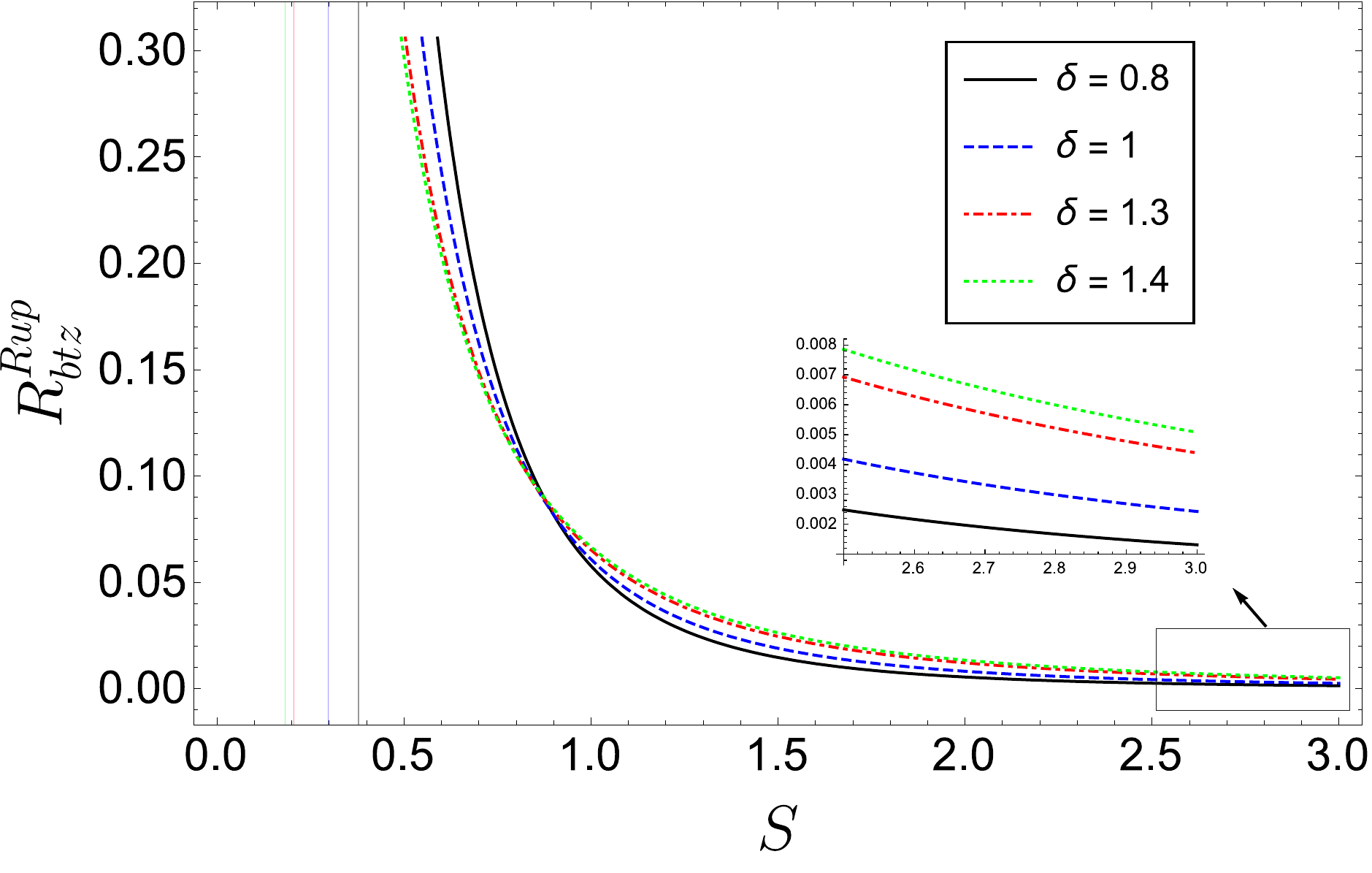}
\caption{Plot of $R^{(Rup)}_{btz}$ as a function of $S$ for various values of $\delta$ and fixed $P=0.1$ (online colors).}
\label{Fig8}
\end{center}
\end{figure}

Following the same strategy as in the previous subsection, 
Tsallis entropy effects on  
microinteractions can be studied by computing Ruppeiner curvature, for instance, 
on the $S-P$ plane. The explicit expression of $R^{Rup}_{btz}$
is given below 
\be
R^{Rup}_{btz}(S,P)=\frac{128\hspace{0.2mm} \pi Q^2 P_{btz} S_{btz}^{\frac{2-\delta}{\delta}}\left(2+\delta\right)}{\left(256 \hspace{0.2mm} P_{btz}S_{btz}^{\frac{2}{\delta}}+\pi \delta Q^2 \right)^2}\,,
\ee
which correctly recovers the result of~\cite{Ghosh:2020kba,Xu:2020ftx} for 
$\eta\rightarrow0$ and $\delta\rightarrow1$.

\begin{figure}[t]
\begin{center}
\includegraphics[width=10cm]{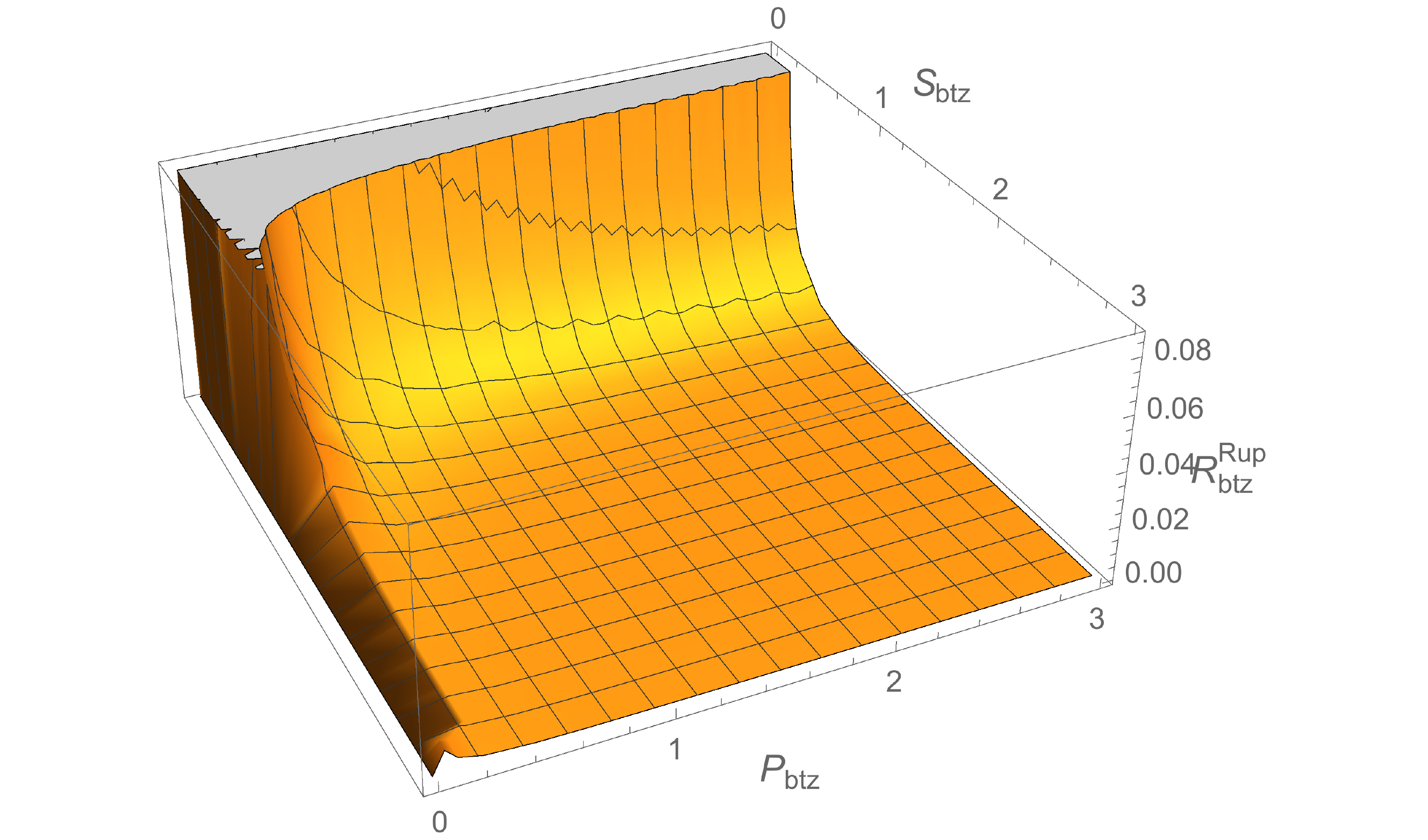}
\caption{3D plot of $R^{(Rup)}_{btz}$ as a function of $S$, $P$ and fixed $\delta=1.3$ (online colors).}
\label{Rup3d}
\end{center}
\end{figure}

The behavior of $R^{Rup}_{btz}$ as a function of $S$ and for various values of $\delta$ is displayed in Fig.~\ref{Fig8}.   
As discussed for AdS BHs,
we observe that $R^{Rup}_{btz}>0$ and vanishes asymptotically, 
with Tsallis entropy affecting
the rate of decrease. Although initially the $\delta<1$ ($\delta>1$) curve seems to decrease slower (faster) than the $\delta=1$ one,
this tendency is reversed as $S$ increases, in agreement with the previous considerations  on  AdS BHs. In Fig.~\ref{Rup3d}
we show the 3D plot of  $R^{Rup}_{btz}$ versus $S$ and $P$ for fixed $\delta=1.3$.

 Non-negativity of $R^{Rup}_{btz}$ is consistent with results in recent literature (see, for instance~\cite{Ghosh:2020kba}), where it has been recorded that microinteractions are always repulsive for normal BTZ BHs carrying charge and/or angular momentum. However, different features might be exhibited by exotic BTZ BHs, where both attraction and repulsion dominated regions do exist. Also, we remark that a behavior similar to that in Fig.~\ref{Fig8} has been found in~\cite{Jawad:2022lww} in the context of BH thermodynamics based on Barrow entropy~\cite{Barrow:2020tzx}, which is a generalization of Bekenstein-Hawking formula motivated
by quantum gravitational considerations (see also~\cite{Saridakis:2020zol,Leon:2021wyx,Sheykhi:2021fwh,Sheykhi:2023,Luciano:2022pzg,Sheykhi:2022gzb,Luciano:2022ffn,Luciano:2023wtx,Cimdiker:2022ics,Luciano:2023roh} for other recent applications of Barrow entropy in BH physics and cosmology). Thus repulsion among BH microstructures appear
as a rather general result, which is not spoilt by either
global monopole or deformed entropy effects. 

Finally, we emphasize that analogous considerations can be
carried out in the coordinate spaces $S-V$, $T-V$ and $T-P$. 
As a matter of example, we consider the $S-V$ space, 
where the temperature and thermodynamic pressure take the form
\begin{eqnarray}
T_{btz}&=&\frac{\pi Q^2 V_{btz}}
{16\delta S_{btz} \left(4S_{btz}^{\frac{2}{\delta}}-\pi V_{btz}\right)}\,, \\[2mm]
P_{btz}&=&\frac{\pi Q^2}{32\left(4S_{btz}^{\frac{2}{\delta}}-\pi V_{btz}\right)}\,,
\end{eqnarray}
respectively. Following~\cite{Xu:2020ftx}, we can obtain the expression of thermodynamic curvature as
\be
R^{Rup}_{btz}(S,V)=\frac{4\left(2+\delta\right)S_{btz}^{\frac{2-\delta}{\delta}}\left(4S_{btz}^{\frac{2}{\delta}}-\pi V_{btz}\right)}{\left[\pi \delta V_{btz}-4\left(2+\delta\right)S_{btz}^{\frac{2}{\delta}}\right]^2}\,.
\ee
It is straightforward to check the identity 
\be
R^{Rup}_{btz}(S,P)=R^{Rup}_{btz}(S,V)\,,
\ee
(and similarly for $R^{Rup}_{btz}(T,V)$ and $R^{Rup}_{btz}(T,P)$), 
thus generalizing the result obtained in~\cite{Xu:2020ftx} to Tsallis framework too. 
Further comments on the above results are discussed in the concluding section. 

\section{Conclusions and Outlook}
\label{Conc}
We have analyzed Tsallis entropy-based geometrothermodynamics and phase transitions of charged AdS BHs with global monopole in an extended phase space. We have treated the cosmological constant and its conjugate quantity as thermodynamic variables related to the pressure and volume, respectively. The equation of state has been studied by employing the standard thermodynamic techniques. We have shown that BHs exhibit a first-order small-large phase transition that is similar to the liquid-gas change of van der Waals fluids. While not affecting the qualitative behavior of $P-v$ diagrams and swallow tail of Gibbs free energy, Tsallis entropy has non-trivial impact on the critical volume, temperature and pressure. In spite of this, we can still recognize a law of corresponding states-like equation. The latter is correctly recovered in the $\delta=1$ limit.
To further explore the role of Tsallis entropy in the
BH/van der Waals fluid correspondence, we have
computed the basic critical exponents by expanding the generalized
law of corresponding states around the critical point of phase
transition. Remarkably, these exponents remain unaffected,
which endorse our comparative picture also at quantitative level.

Finally, we have analyzed BH geometric thermodynamics within
Ruppeiner framework on the $S-P$ plane, which allows to establish the nature of of microscopic forces among BH constituents by looking at the sign of the scalar curvature. 
This study has revealed that
AdS BHs exhibit  vanishing/prevailing repulsive interactions at different phases of entropy.   In particular, we have shown that non-additivity influences the intervals where net interactions become manifest/disappear. A possible explanation for this result has been furnished based on the behavior of the two fluid model, where the nature of prevailing interactions is fixed by the relative number densities of the molecules of the two fluids.
We have then extended geometrothermodynamic analysis to
$(2 +1)$-dimensional charged BTZ BHs. Following~\cite{Jawad:2022lww}, we have derived the entropy-pressure and entropy-volume corrected curvatures, which still indicate
purely repulsive forces among microscopic structures. This is in line with recent results in literature~\cite{Cai:1998ep,Ghosh:2020kba}.

Further aspects are yet to be investigated. First, we remark
that a comprehensive study of BH geometrothermodynamics should involve an ab initio derivation of modified field equations based on Tsallis entropy. This will be our next step.
In particular, it is suggestive to understand whether and, if so, how the present formalism departs from a theory of gravity fully
compatible with non-additive Tsallis prescription.
Furthermore, we aim at generalizing the above study to the case of rotating BHs, which have interesting features. Indeed, it is known that at thermodynamic level, the angular momentum of BHs behaves like an electric charge, thus originating peculiar repulsive interactions. This poses the problem of how such effects are influenced by Tsallis entropy. Along this line, similar considerations can be applied to exotic BTZ BHs, which have been shown to exhibit a rich phenomenology with both attraction and repulsion dominated regions.

On the other hand, possible further perspectives concern
the investigation of BH critical phenomena in other deformed statistics, such as R\'eny~\cite{Reny} and Kaniadakis~\cite{Kania1} entropy-based statistics, which arise in quantum information and relativity theory, respectively (see~\cite{Kania2} for a recent review on gravitational and cosmological applications of Kaniadakis entropy). Also, one can conduct a similar analysis for the case of 
loop quantum gravity~\cite{LQG1,LQG2,LQG3} 
and Barrow~\cite{Barrow:2020tzx,Ghaffari:2022skp} entropy, the latter being an entropic  form that emerges in quantum gravity but has no statistical roots. In this scenario, the entropy-area law is modified according
\be
S_{BH}\,\,\longrightarrow\,\, S_{Barrow}=\left(S_{BH}\right)^{1+\frac{\Delta}{2}}\,,
\ee
where the deformation parameter $0\le\Delta\le1$ quantifies
quantum gravitational corrections, so that $\Delta=1$
leads to the maximal deformation, while $\Delta=0$ 
yields the standard horizon entropy. Though Barrow entropy
arises from different physical motivations,  in its essence
it is equivalent to Tsallis entropy~\eqref{TsaEN}
under the reparameterization $\delta\rightarrow1+\Delta/2$. 
This implies that all the discussions of thermodynamic quantities
and phase transitions of BHs are expected to be the same
as the ones presented here, though in a restricted range of parameters consistent with the observations below Eq.~\eqref{Volcrit}.

Finally, we mention that recent works have explored
the impact of quantum gravity-induced modifications of Heisenberg
principle (Generalized Uncertainty Principle, GUP) on non-extensive BH thermodynamics~\cite{Alonso,Cimidiker:2023kle}. In this context, the challenge is to relate the present results with those
obtained in the GUP framework to better understand the
interplay between deformed uncertainty relations
and non-Gaussian statistics. Preliminary insights in this direction
appear in~\cite{Shababi:2020evc,Luciano:2021ndh}, where
this connection is ascribed to the existence of a GUP minimal length that modifies the phase space structure and the emergent statistics, and in~\cite{Jizba:2022icu}, which unveils that coherent states associated to GUP coincide with Tsallis probability amplitude. Work along these directions is still in progress and will be presented elsewhere.

\acknowledgements
G. G. L. acknowledges the Spanish ``Ministerio de Universidades''
for the awarded Maria Zambrano fellowship and funding received
from the European Union - NextGenerationEU.
He is also grateful for participation to the COST
Action CA18108  ``Quantum Gravity Phenomenology in the Multimessenger Approach''  and LISA Cosmology Working group.
Finally, he acknowledges S. Soroushfar and L. Petruzziello for useful discussions.

\end{document}